\documentclass[sigconf]{acmart}
\usepackage{epsfig,endnotes}
\usepackage[normalem]{ulem}
\usepackage{graphicx}
\usepackage{amsmath}
\usepackage{mathtools}
\usepackage{url}
\usepackage{hyperref}
\usepackage{cleveref} 

\usepackage{tikz,tikz-3dplot}
\usetikzlibrary{arrows,shapes,positioning,shadows,trees,automata,graphs,decorations.pathmorphing,shadows.blur}
\usepackage{adjustbox}
\usepackage{amssymb}
\usepackage{amsthm}
\usepackage{listings}
\usepackage{color}
\usepackage{subcaption}
\usepackage{booktabs}

\usepackage{enumitem}

\usepackage{flushend}

\makeatletter
\renewcommand\subsubsection{\@startsection{subsubsection}{3}{10pt}%
  {-.5\baselineskip \@plus -2\p@ \@minus -.2\p@}%
  {-3.5\p@}%
  {\@subsubsecfont\normalfont\bfseries\@adddotafter}}
\makeatother

\Crefname{example}{Example}{Examples}
\crefname{example}{example}{examples}

\usepackage{array}
\newcolumntype{L}[1]{>{\raggedright\let\newline\\\arraybackslash\hspace{0pt}}p{#1}}
\newcolumntype{C}[1]{>{\centering\let\newline\\\arraybackslash\hspace{0pt}}p{#1}}
\newcolumntype{R}[1]{>{\raggedleft\let\newline\\\arraybackslash\hspace{0pt}}p{#1}}

\definecolor{greenet}{RGB}{0,160,0}
\definecolor{bluet}{RGB}{0,170,255}
\definecolor{oranget}{RGB}{247,51,0}

\tikzset{
    basic/.style      = {draw, font=\small\ttfamily, , fill=white},
    empty/.style      = {draw=none},  
    label/.style      = {font=\small\sffamily, text centered},
    node/.style       = {basic, rounded corners=2pt, thick, align=center, fill=white, draw=black, minimum size=0.6cm},
    state/.style      = {basic, circle,                         align=center, draw=gray, minimum size=0.6cm},
    statetran/.style  = {basic, rectangle, rounded corners=2pt, align=center, draw=black, minimum size=0.6cm},
    statesym/.style   = {basic, rectangle, rounded corners=2pt, align=center, draw=black, minimum size=0.6cm},
    evt/.style        = {basic,                                 align=center, draw=black, minimum size=0.6cm},
    evtnet/.style     = {evt, draw=greenet},
    evtcall/.style    = {evt, draw=bluet},
    prop/.style       = {basic, dashed, draw=black},
    var/.style        = {basic, rounded corners=2pt, text=black, draw=gray},
    intra/.style      = {->, >=stealth',draw=gray},
    toempty/.style    = {intra, densely dotted},
    inter/.style      = {->, >=stealth', color=gray, densely dotted},
    causes/.style     = {inter, color=bluet},
    parses/.style     = {inter, color=greenet},
    accepts/.style    = {inter, color=oranget},
    propag/.style     = {inter, ->, color=purple},
    abstracts/.style  = {inter, ->, color=red},
    ext/.style        = {inter, ->, color=gray},
    bel/.style        = {inter, color=orange},
    wavenodec/.style  = {decoration={snake, post length=0.5mm, pre length=0.5mm, amplitude=0.5mm, segment length=2mm}},
    wave/.style       = {decorate, wavenodec},
    layer/.style      = {shade, blur shadow={shadow scale=1, shadow xshift=0.5mm, shadow yshift=-0.5mm, fill=gray}, top color=white, densely dotted},
    ptnode/.style     = {shape=rectangle, rounded corners, draw, align=center, draw=gray, fill=white},
    root/.style       = {ptnode, font=\itshape},
    term/.style       = {ptnode, font=\ttfamily},
    nterm/.style      = {ptnode, font=\itshape}
}

\lstdefinestyle{sql}{ %
  basicstyle=\footnotesize\ttfamily\bfseries,        
  breakatwhitespace=false,         
  breaklines=true,                 
  captionpos=b,                    
  deletekeywords={...},            
  escapeinside={\%*}{*)},          
  extendedchars=true,              
  frame=single,                    
  rulecolor=\color{lightgray},
  keywordstyle=\textcolor{blue},       
  keywords={UPDATE, SET, WHERE},        
  keepspaces=true,                 
  showspaces=false,                
  showstringspaces=false,          
  showtabs=false,                  
  stepnumber=2,                    
  tabsize=2,                       
  title=\lstname                   
}

\lstdefinestyle{http}{ %
  basicstyle=\footnotesize\ttfamily\bfseries,        
  breakatwhitespace=false,         
  breaklines=true,                 
  captionpos=b,                    
  deletekeywords={...},            
  escapeinside={\%*}{*)},          
  extendedchars=true,              
  frame=single,                    
  rulecolor=\color{lightgray},
  keywordstyle=\textcolor{blue},       
  keywords={GET, POST, HTTP},        
  keepspaces=true,                 
  showspaces=false,                
  showstringspaces=false,          
  showtabs=false,                  
  stepnumber=2,                    
  tabsize=2,                       
  title=\lstname                   
}

\newcommand{\empirical}[1]{#1}
\newcommand{\gp}[1]{}
\newcommand{\cro}[1]{}
\newcommand{\mj}[1]{}
\newcommand{\mb}[1]{}

\newcommand{\tool}[0]{\texttt{Deemon}}

\crefformat{footnote}{#2\footnotemark[#1]#3}

\theoremstyle{definition}
\newtheorem{defn}{Definition}

\lstdefinestyle{custompy}{
  belowcaptionskip=1\baselineskip,
  breaklines=true,
  language=Python,
  showstringspaces=false,
  basicstyle=\small\ttfamily,
  keywordstyle=\bfseries\color{green!40!black},
  commentstyle=\itshape\color{purple!40!black},
  identifierstyle=\color{blue},
  stringstyle=\color{orange},
  frame=single,
  numbers=left
}

\newcommand{\acsrf}[0]{aCSRF}

\newcommand{\nodelabel}[1]{\mathtt{#1}}
\newcommand{\edgelabel}[1]{\mathtt{#1}}
\newcommand{\propname}[1]{\mathtt{#1}}
\newcommand{\propvalue}[1]{\mathtt{#1}}

\newcommand{\eqdef}[0]{\stackrel{def}{\mathtt{:=}}}

\newcommand{\Q}[1]{\mathtt{#1}}

\newcommand{\QLabel}{\Q{Label}}

\newcommand{\QAbs}{\Q{Abs}}
\newcommand{\QParses}{\Q{Parses}}
\newcommand{\QHTTPReq}{\Q{HTTPReq}}

\newcommand{\QState}{\Q{State}}

\newcommand{\QTrans}{\Q{Trans}}

\newcommand{\QFSMAccepts}{\Q{Accepts}}

\newcommand{\QCaused}{\Q{Causes}}

\newcommand{\QSC}{\Q{SC}}

\begin{document}

\title{\tool{}: Detecting CSRF with Dynamic Analysis and Property Graphs}

\author{Giancarlo Pellegrino}
\affiliation{
    \institution{CISPA, Saarland University\\ Saarland Informatics Campus}
    \streetaddress{}
    \city{} 
    \state{} 
    \postcode{}
}
\email{gpellegrino@cispa.saarland}

\author{Martin Johns}
\affiliation{
    \institution{SAP SE}
    \streetaddress{}
    \city{} 
    \state{} 
    \postcode{}
}
\email{martin.johns@sap.com}

\author{Simon Koch}
\affiliation{
    \institution{CISPA, Saarland University\\ Saarland Informatics Campus}
    \streetaddress{}
    \city{} 
    \state{} 
    \postcode{}
}
\email{s9sikoch@stud.uni-saarland.de}

\author{Michael Backes}
\affiliation{
    \institution{CISPA, Saarland University\\ Saarland Informatics Campus}
    \streetaddress{}
    \city{} 
    \state{} 
    \postcode{}
}
\email{backes@cispa.saarland}

\author{Christian Rossow}
\affiliation{
    \institution{CISPA, Saarland University\\ Saarland Informatics Campus}
    \streetaddress{}
    \city{} 
    \state{} 
    \postcode{}
}
\email{rossow@cispa.saarland}

\renewcommand\footnotetextcopyrightpermission[1]{}




\begin{abstract}

Cross-Site Request Forgery (CSRF) vulnerabilities are a severe class of web
vulnerabilities that have received only marginal attention from the research
and security testing communities. While much effort has been spent on
countermeasures and detection of XSS and SQLi, to date, the detection of CSRF
vulnerabilities is still performed predominantly manually.

In this paper, we present \tool{}, to the best of our knowledge the first
automated security testing framework to discover CSRF vulnerabilities. Our
approach is based on a new modeling paradigm which captures multiple aspects
of web applications, including execution traces, data flows, and architecture
tiers in a unified, comprehensive property graph. We present the paradigm and
show how a concrete model can be built automatically using dynamic traces.
Then, using graph traversals, we mine for potentially vulnerable operations.
Using the information captured in the model, our approach then automatically
creates and conducts security tests, to practically validate the found CSRF
issues. We evaluate the effectiveness of \tool{} with 10 popular
open source web applications. Our experiments uncovered \empirical{14}
previously unknown CSRF vulnerabilities that can be exploited, for instance,
to take over user accounts or entire websites.

\end{abstract}

\maketitle

\section{Introduction}

No other vulnerability class illustrates the fundamental flaws of the web
platform better than Cross-Site Request Forgery (CSRF): Even a brief visit
to an untrusted website can cause the victim's browser to perform
authenticated, security-sensitive operations at an unrelated, vulnerable web
application, without the victim's awareness or consent. To achieve this, it is
sufficient to create a single cross-origin HTTP request from the attacker
webpage, a capability that is native to the Web ever since Marc Andreessen
introduced the {\tt img} HTML tag element in February
1993~\cite{img-tag-intro}.

Since its discovery in 2001~\cite{Schreiber_ses-riding}, CSRF vulnerabilities
have been continuosly ranked as one of the top three security risks for web
applications, along with cross-site scripting (XSS) and SQL injection
(SQLi)~\cite{owasp_top10s,Barth:2008:RDC:1455770.1455782,johnny}. Successful CSRF
exploitations can result in illicit money
transfers~\cite{Zeller_cross-siterequest}, user account takeover~\cite{Sudhodanan2017},
 or remote server-side command
execution~\cite{deepsec-2007}, to name only a few publicly documented cases.
In the past, similar vulnerabilities have been discovered in many popular
websites including Gmail~\cite{gmail_csrf}, Netflix~\cite{netflix-csrf}, ING
Direct~\cite{Zeller_cross-siterequest}, and, more recently, in Google, Skype,
and Ali Express websites~\cite{Sudhodanan2017}.

Despite its popularity, CSRF has received only marginal attention, compared to
SQLi and XSS. Most of the previous efforts have been spent in proposing
active~\cite{JovanovicKK06, Johns_requestrodeo:client, Kerschbaum_CSRF} or
passive~\cite{Barth:2008:RDC:1455770.1455782} defense mechanisms, and little
has been done to provide developers and practitioners with effective
techniques to \emph{detect} this class of vulnerabilities. Classical
vulnerability detection techniques utilize dynamic~\cite{stateaware, authscan,
DBLP:conf/ndss/PellegrinoB14, Pellegrino2015} and static analysis
techniques~\cite{webssari, Dahse_stored, MACE, whitebox_logic, backes2017},
while mainly focusing on injection vulnerabilities~\cite{stateaware, webssari,
Dahse_stored} or flaws specific to the application logic
layer~\cite{stateaware, DBLP:conf/ndss/PellegrinoB14, MACE, whitebox_logic}.
Unfortunately, none of the existing techniques are easily applicable to CSRF.
As a result, to date, CSRF vulnerabilities are still predominately discovered
by manual inspection~\cite{Sudhodanan2017}.

\vspace{3pt}\noindent\textbf{Our Approach}---We take a step
forward by presenting \tool{}, a model-based security testing framework to
enable the detection of CSRF vulnerabilities. To the best of our knowledge,
this is the first automated technique that targets the detection of CSRF.
\tool{} automatically augments the execution environment of a web application,
to enable the unsupervised generation of dynamic {\it execution traces}, in
the form of, e.g., network interaction, server-side execution, and database
operations. Using these traces, \tool{} infers a \emph{property graph}-based
model of the web application capturing different aspects such as state
transitions and data flow models in a unified representation. Operating on the
resulting model, \tool{} uses graph traversals to identify security-relevant
state-changing HTTP requests, which represent CSRF vulnerability candidates.
Finally, leveraging the augmented application runtime, \tool{} validates the
candidate's vulnerability against the real web applications.

We assessed \tool{} against 10 popular open source web applications and
discovered \empirical{14} previously-unkown CSRF vulnerabilities in four of
them. These vulnerabilities can be exploited to take over websites, user
accounts, and compromise the integrity of a database. Finally, we analyzed our
test results to assess the current awareness level of the CSRF
vulnerabilities. In two cases, we identified alarming behaviors in which
security-sensitive operations are protected in a too-selective manner.

To summarize, we make the following contributions:

\begin{itemize}
	
	\item We present \tool{}, an automated, dynamic analysis, security testing
	technique to detect CSRF vulnerabilities in productive web applications;

	\item We present a new modeling paradigm based on \emph{property graphs},
	that is at the core of \tool{};
	
	\item We show how \tool{}'s models can be instantiated in an unsupervised,
	automatic fashion, requiring only selected GUI interaction recordings;
	
	\item We report on a practical evaluation of \tool{} using 10 popular web
	applications, which uncovered \empirical{14} severe CSRF vulnerabilities; and

	\item We assess the CSRF awareness level and discover alarming
	behaviors in which security-sensitive operations are protected in a
	selective manner.

\end{itemize}


\section{Cross-site Request Forgery (CSRF)}
\label{sec:background}
\label{sec:attack}

\begin{figure}[t]
  \centering
  \includegraphics[width=\columnwidth]{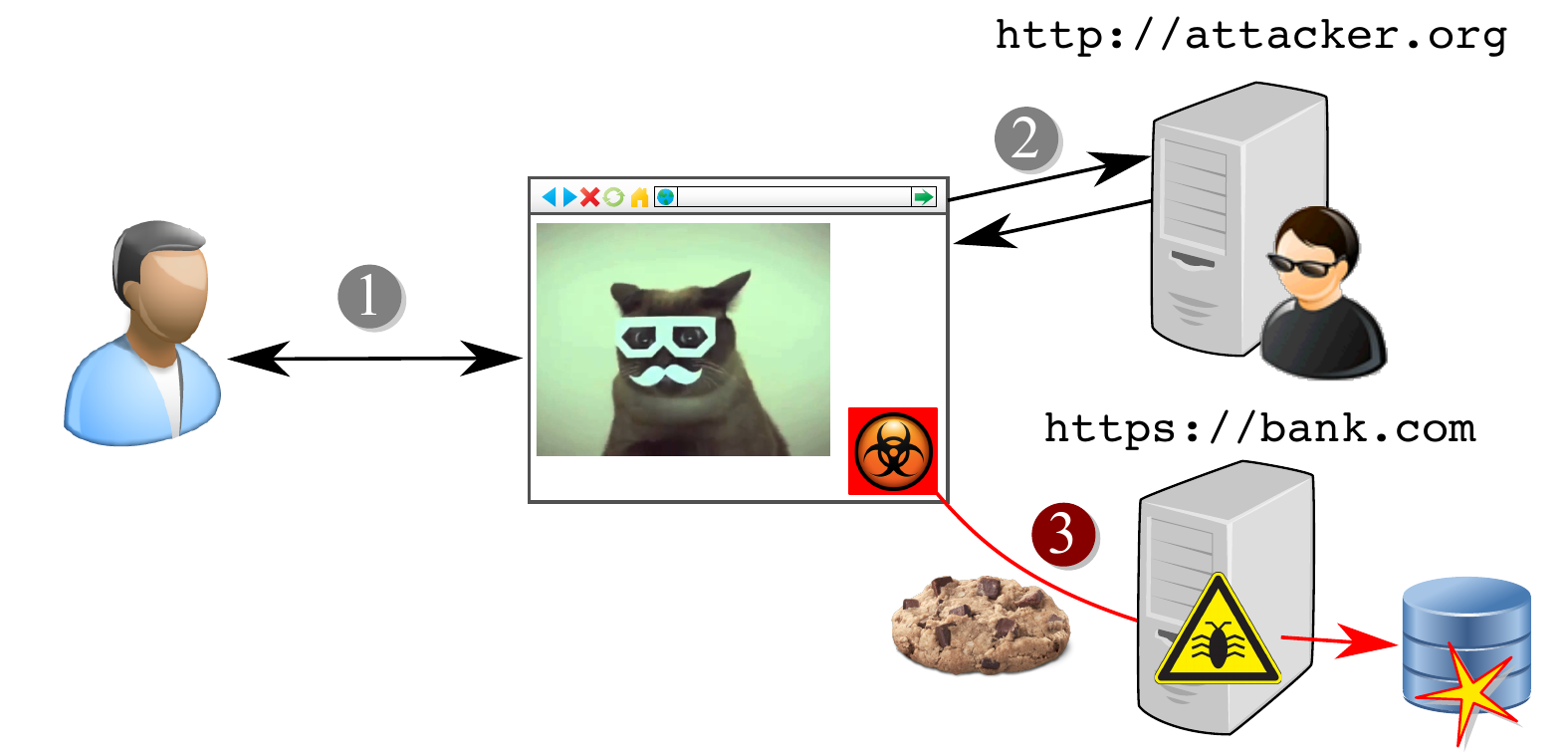}
  \caption{Authenticated CSRF attack.}
  \label{fig:CSRF example}
  \label{fig:csrf}
\end{figure}

In CSRF attacks, an attacker tricks the web browser of the victim to send a
request to a vulnerable honest website in order to cause a desired,
security-sensitive action, without the victim's awareness or consent. Desired
actions can be, for example, illicit money
transfers~\cite{Zeller_cross-siterequest}, resetting account
usernames~\cite{Sudhodanan2017}, or the execution of specific server-side
commands~\cite{deepsec-2007}. CSRF attacks can be distinguished into two main
categories: authenticated and login CSRF. In an \emph{authenticated} CSRF
(\acsrf{}), a pre-established, authenticated user session between the victim's
web browser and the targeted web application exists. In a \emph{login} CSRF,
such a relationship does not exist, but the goal of the attacker is to log the
victim in by using the attacker's credentials. In the remainder of this paper,
we focus on \acsrf{} attacks, the significantly larger category. An extensive
overview of login CSRF is provided by Sudhodanan et al.~\cite{Sudhodanan2017}.

\Cref{fig:csrf} shows an example of an \acsrf{} attack. The actors of an
\acsrf{} attack are the user (i.e., the victim), a vulnerable target website (e.g.,
\texttt{bank.com}, a home banking website), and an attacker controlling a
website (e.g., \texttt{attacker.org}). In an \acsrf{} attack, the victim is
already authenticated with the target website. Upon a successful
authentication, the website of the bank persists an authenticated session
cookie in the user's web browser. From this point on, whenever the user visits
the website of the bank, the browser includes this session
cookie~\cite{rfc6454}. An attacker can exploit this behavior of the browser as
follows. First, she prepares an HTML page containing malicious code. The goal
of this code is to perform a \emph{cross-origin} HTTP request to the website
of the bank. This can be implemented in different ways, e.g., with an HTML
\texttt{iframe} tag, a hidden HTML form with self-submitting JavaScript code,
or via the XMLHttpRequest JavaScript API~\cite{ajax}. Then, when a victim
visits the malicious page, her browser generates such a request, which
automatically includes the the session cookie. The bank checks the cookie, and
executes the required operation. If the HTTP request encodes, e.g., a request
to update user password, then the bank executes it without the actual consent
of the bank account owner.

More formally, we define an \acsrf{} vulnerability as follows.

\begin{defn}

A web application (e.g., bank.com) exposes an \acsrf{}
vulnerability, if the web application accepts an HTTP request (e.g., message 3)
with the following properties:

\begin{description}
  
  \item[(P1)] The incoming request causes a security-relevant state
  change of the web application.
  
  \item[(P2)] The request can be reliably created by an attacker, i.e., the
  attacker knows all the required parameters and values of the request.
  
  \item[(P3)] The request is processed within a valid authentication context of a
  user.

\end{description}
\end{defn}

Cross-origin requests can be used in other attacks without necessarily causing
a server-side state transition, e.g., accessing user data stored in the target
website. These attacks are addressed by the \emph{same-origin policy}
(SOP)~\cite{rfc6454} for cross-origin requests, which blocks the access to HTTP
responses. However, the SOP does \emph{not} prevent the browser from
performing HTTP requests. To defend against malicious cross-origin requests,
the server-side program can check the request origin via the header
\texttt{Origin}. However, this header may not be present in a request. The
current best-practice
\acsrf{} protection is the so-called \emph{anti-CSRF
token}~\cite{Barth:2008:RDC:1455770.1455782}. An anti-CSRF token is a
pseudo-random value that is created by the server and explicitly integrated
into the request by the client. Various methods exist to implement anti-CSRF
tokens, including hidden form fields or custom HTTP headers. Further
implementation details are left out of this document for brevity.


\section{Challenges in Detecting aCSRF}
\label{sec:challenges}

A security testing approach designed to detect \acsrf{} vulnerabilities faces
two distinct classes of challenges, neither of them met by the current
state-of-the-art in security testing: \textit{detection challenges} and
\textit{operational challenges}, as discussed next.

\subsection{Detection Challenges} 

Detecting aCSRF requires reasoning over the relationship between the
application state, the roles and status of request parameters, and the
observed sequences of state transitions. This leads to a set of specific
detection challenges that directly result from the unique characteristics of
the vulnerability class.

\vspace{4pt}\noindent\textbf{(C1) State Transitions}---The first challenge is
to determine when a state transition occurs. Server-side programs implement
several operations; not all of them affect the state of the application.
Consider, for instance, the function of searching for a product in an online
store: The user provides search criteria, causing the server-side program to
search its database for matching products. The permanent state of the user's
data in the application is unaffected by this process. However, other
operations change the state of the program. Consider a user that wants to
change their login password. The server-side program uses the new password to
update the database entry. From that point on, the old password is no longer
accepted; thus, the state has changed.

Existing tools such as \emph{web application scanners} (See,
e.g.,~\cite{Kals:2006:SWV:1135777.1135817,johnny}) mainly operate in a
black-box manner. They crawl a web application and send requests with crafted
input. Vulnerabilities are detected by inspecting responses. This approach
works well with XSS and SQLi, but does not scale to CSRF as it cannot discern
when a request changes the server-side state. Web crawlers can be made aware
of server-side states by inferring a model capturing transitions via webpage
comparisons: If the HTML content is similar, then they originate from the same
state (See, e.g., Doup\'{e} et al.~\cite{stateaware}). However, as pages
contain dynamic content, the similarity may not be determined precisely, thus
resulting in inaccurate models. Finally, techniques to infer models are often
specific to the function being tested (See,
e.g.,~\cite{Wang:2012:SMY:2310656.2310691, DBLP:conf/ndss/PellegrinoB14}).
\acsrf{} vulnerabilities can affect any function of a web application; thus,
function-specific models cannot be easily used to detect \acsrf{}
vulnerabilities.

\vspace{4pt}\noindent\textbf{(C2) Security-Relevant State Changes}---The
second challenge is to determine the relevance of a state transition. State
transitions can be the result of operations such as event logging and tracing
user activity. These operations indeed change the state of the server, but
they are not necessarily security relevant. While a human may distinguish the
two cases, automated tools without a proper description of the application
logic may not tell the two transitions apart. Especially for static analysis
approaches, security-neutral state changes are indistinguishable from aCSRF
candidates.

\vspace{4pt}\noindent\textbf{(C3) Relationships of Request Parameters and
State Transitions}---The third challenge consists in determining the relations
between request parameters and state transitions. The identification of these
relations is relevant for the detection of \acsrf{} vulnerabilities. For
example, consider a parameter carrying a random security token. An attacker
may not be able to guess such a parameter, thus preventing her from
reconstructing the HTTP request. The identification of these parameters is
important, as it suggests the presence of anti-CSRF countermeasures, and can
be used to develop a testing strategy. For example, the tester may replay the
request without the token to verify whether the web application properly
enforces the use of the security token. Another example is a parameter
carrying a user input, e.g., a new user password, that is stored in the
database. An attacker can use this parameter to hijack a user account by using
a password that she controls.

Existing techniques do not determine the relations between parameters and
state transitions. Web scanners attempt to identify security tokens by
matching parameter names against a predefined list of patterns, e.g., the
parameter being called \texttt{token}. In general, to determine the role of a
request parameter, we need to determine the type of relations with state
transitions. As these parameter values traverse the tiers of an application,
we may need to track their flow across all tiers, e.g., presentation, logic,
and data. The resulting model of data flows can be enriched with type
information, e.g., both semantic and syntactic types, to determine the nature
of the value, e.g., user-controlled or pseudo-random.

\subsection{Operational Challenges}

The operational challenges in detecting aCSRF are direct consequences of
addressing the \textit{detection challenges} in the context of dynamic
security testing.

\vspace{4pt}\noindent\textbf{(C4) Transitions in Non-Trivial Application
Workflows}---The fourth challenge is to reach state-changing requests in
non-trivial web application workflows. Dynamic analysis techniques such as
unsupervised web scanners explore HTML webpages using breadth- or depth-first
search algorithms. However, these algorithms are too simplistic to cope with
the complexity of modern web application workflows in which users need to
perform a specific sequence of actions. Likewise, static analysis techniques
look for patterns in the source code to determine the presence of a
vulnerability. However, without a proper description of the workflow, static
approaches scale poorly to large applications.

\vspace{4pt}\noindent\textbf{(C5) Side-Effect-Free Testing}---Dynamic testing
for aCSRF vulnerabilities is centered around the iterative detection of
state-chang-ing HTTP requests (Challenges C1 \& C2). However, as such requests
indeed \textit{change} the application state, all further test requests
attempting to assess the relationships of request parameters and state
transitions (C3) will most likely operate on a now-invalid state. Take for
example the dynamic testing for aCSRF vulnerabilities in a shopping cart web
application. As soon as a test request has submitted the cart beyond the
check-out state, no further security testing on this state transition can be
conducted, as the active shopping cart ceases to exist. Thus, a testing method
is needed, that allows evaluation of HTTP request-induced state changes in a
side-effect-free manner.\gp{I think for this one it would be great to have
also references on the challenges of resetting the testbed to a known state}

\vspace{4pt}\noindent\textbf{(C6) Comprehensive, Reusable Representation of
Application Functionality}---The final challenge results from the previous
challenges. To detect security-relevant state changes, we need to combine
aspects of the web application. On the one hand, we have transitions
describing the evolution of the internal states of the server-side program. On
the other hand, we have data flow information capturing the propagation of
data items across tiers and states. These aspects can be represented by means
of \emph{models}.

In literature, there are many languages and representations to specify models,
ranging from formal languages~\cite{Hierons:2009:UFS:1459352.1459354} to
custom models tailored to the specific application function being tested
(e.g.,~\cite{Wang:2012:SMY:2310656.2310691, DBLP:conf/ndss/PellegrinoB14}).
Often, the combination of models has been addressed in a custom way. The
shortcoming of this approach is that the combination is achieved without
specifying the relationships between the models, thus making it hard to reuse
it for other techniques. Another approach is to create representations that
combine elements of individual models, such as extended finite-state machines
that fire transitions when certain input conditions
hold~\cite{Hierons:2009:UFS:1459352.1459354}. However, defining new modeling
languages may not scale well, as a new language is required as soon as new
aspects need to be included.


\begin{figure*}[t]
  \centering
  \includegraphics[width=\textwidth]{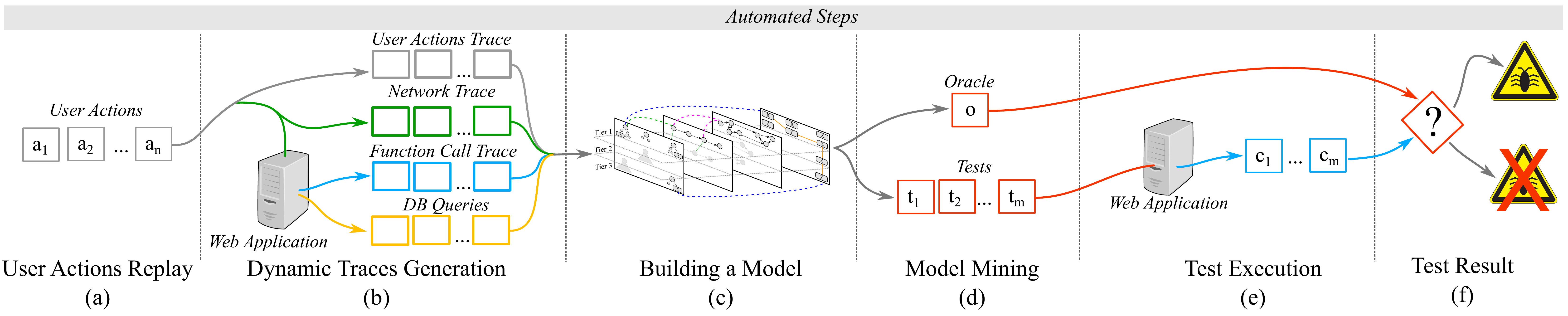}
  \caption{Overview of the detection phase of \tool{}.}
  \label{fig:overview}
\end{figure*}

\section{\tool{}: Overview}
\label{sec:our_approach}

To overcome the challenges of \Cref{sec:challenges}, we developed
\tool{}\footnote{Source code and documentation of \tool{} can be downloaded
here \url{https://github.com/tgianko/deemon}}, an application-agnostic,
automated framework designed to be used by developers and security analysts
during the security testing phase of the software development life-cycle. The
current version of \tool{} supports PHP-based web applications that use MySQL
databases, and it can be easily extended to support other languages and
databases. The key features of \tool{} that allow for addressing our
challenges are the following:

\begin{itemize}

    \item \tool{} infers models from program execution observations capturing
    state transitions and data flow information (Challenges C1 \& C3).

    \item \tool{} uses \emph{property graphs} to represent these models. This
    provides a uniform and reusable representation and defines precise
    relationships between models by the means of \emph{labeled edges}
    (Challenge C6).

    \item \tool{} leverages a programmatic access to the property graph via
    \emph{graph traversals} to identify security-relevant state changes
    (Challenge C2).

    \item \tool{} augments the execution environment of a web application and then
    reproduces a set of user actions to observe server-side program execution
    (Challenge C4).

    \item \tool{} relies on virtualized environments to test web applications.
    This enables full control of the web application by taking and restoring
    \emph{snapshots} (Challenge C5);

\end{itemize}

\tool{} takes as input a set of user actions and an application container of
the web application under test. \tool{} operates in phases:
\emph{instrumentation} and \emph{detection}. In the first phase,
\tool{} modifies the application container to insert sensors for the
extraction of network traces, server-side program execution traces, and
sequence of database operations. In the second phase,
\tool{} automatically reproduces user actions, infers a model from the
resulting traces, and tests the web application to detect \acsrf{}
vulnerabilities.

\subsection{Preparation}

\tool{} is meant to support developers and security analysts. In this section,
we briefly present the tool as seen by a user.

\vspace{4pt}\noindent\textbf{Inputs}---The inputs of \tool{} are a set of user
actions and an application container of the web application under test.

\vspace{2pt}\noindent\textit{User Actions}: The first input is a set of user
action sequences (see \Cref{fig:overview}.a) that are provided by the tester.
User actions are artifacts commonly used in security testing~\cite{otgv4} and
there is a plethora of automated tools to create them via web browsers and use
them when testing web applications~\cite{otgv4}. A user action is performed on
the UI of the web application. For example, a user action can be a mouse
click, a key stroke, or an HTML form submission. The sequence of actions
represent a web application functionality. For example, consider the operation
of resetting user credentials. The user actions trace contains the following
actions: \emph{load} \texttt{index.php} page, \emph{click} on change
credential link, \emph{type} new username and password, and \emph{click}
submit. Input traces can also be actions of a privilegded user, e.g., website
administrator, when changing the website configuration from the administrator
panel.

\vspace{2pt}\noindent\textit{Application Container}: The second input of
\tool{} is an application container of the web application under test. An
application container consists of a runtime environment with software,
dependencies and configuration. Web application containers contain the web
application (binary or source code), database server, and application
configuration. Containers are convenient tools as they allow the deployment of
ready-to-use web applications. Nowadays application containers are gaining
momentum and are becoming a popular means to distribute and deploy web
applications.

\vspace{4pt}\noindent\textbf{Outputs}---\tool{} returns a vulnerability report,
listing state-chang-ing HTTP requests that can be used to
perform \acsrf{} attacks.

\subsection{Instrumentation}

Given an application container, \tool{} automatically installs sensors to
monitor the program execution. For example, for PHP-based web applications,
\tool{} adds and enables the \emph{Xdebug}~\cite{xdebug} module of the PHP
interpreter, an extension that generates full function call trees.
Furthermore, \tool{} installs a local HTTP proxy to intercept HTTP messages
exchanged between the server and the browser.

\subsection{Detection}
\label{sec:detection}
The core function of \tool{} is the detection of \acsrf{} vulnerabilities. The
main steps are shown in \Cref{fig:overview} and are all automated. The
detection begins by reproducing the user actions against a running instance of
the web application (\Cref{fig:overview}.a). The sensors installed during the
instrumentation produce execution traces that include network traces and
function call traces (\Cref{fig:overview}.b). \tool{} runs this step twice to
observe, for example, sources of non-determinism such as generation of
pseudo-random data items. Each run is called \emph{session}. From these
traces, \tool{} infers a model which is the composition of simpler models,
e.g., finite-state machine and data flow model with data type information
(\Cref{fig:overview}.c). Then, \tool{} uses model queries to mine both
security tests and an oracle (\Cref{fig:overview}.d), and runs them against
the web application (\Cref{fig:overview}.e). Finally, it evaluates test
results against the oracle to detect CSRF vulnerabilities
(\Cref{fig:overview}.f).

\section{Modeling}
\label{sec:model}

\begin{figure*}[t!]
      \centering
      \begin{adjustbox}{width=1\textwidth}
        \begin{tikzpicture}

\draw[color=gray, densely dotted] (-5.5,-4)  -- (17.2,   -4);
\draw[color=gray, densely dotted] (-1.5,  0.2) -- (-1.5, -8);
\draw[color=gray, densely dotted] (5.5, 0.2) -- ( 5.5, -8);
\draw[color=gray, densely dotted] (13,0.2)   -- (13  , -8);

\node[empty] (DFM) [xshift=-6cm, yshift=-2cm, rotate=90]   {\LARGE{\begin{tabular}{c} Logic Tier \\ (i) \end{tabular}}};
\node[empty] (DFM) [xshift=-6cm, yshift=-6cm, rotate=90]   {\LARGE{\begin{tabular}{c} Data  Tier \\ (ii)\end{tabular}}};

\node[empty] (DFM) [xshift=-3.4cm, yshift=-8.2cm]   {\LARGE{\begin{tabular}{c} Dataflow Models \\ (a) \end{tabular}}};
\node[empty] (DFM) [xshift= 2cm,   yshift=-8.2cm]   {\LARGE{\begin{tabular}{c} Finite-State Machines \\ (b) \end{tabular}}};
\node[empty] (DFM) [xshift= 9.2cm, yshift=-8.2cm]   {\LARGE{\begin{tabular}{c} Parse Trees \\ (c) \end{tabular}}};
\node[empty] (DFM) [xshift=15cm,   yshift=-8.2cm]   {\LARGE{\begin{tabular}{c} Traces \\ (d) \end{tabular}}};

\begin{scope} 

  \begin{scope}[
      xshift=-0.4cm, 
      yshift=-2cm,
      node distance=3cm]
        \node[state] (q1) [xshift=-0.5cm]       {$q_{0}$};
        \node[state] (q2) [right=3cm of q1]   {$q_{1}$};
        \node[state] (q3) [right=1.5cm of q2] {$q_{2}$};

        \node[statetran] (q0_0) [right=0.5cm of q1, text=black]     {$tr(q_{0}, x')=q_{1}$};
        \node[statetran] (q1_1) [above right=0.5cm and 0.2cm of q2] {$tr(q_{1}, x'')=q_{2}$};
        \node[statetran] (q2_0) [below right=0.5cm and 0.2cm of q2] {$tr(q_{1}, x''')=q_{2}$};

        \draw[intra] (q1)   to node [above] {\tiny$\edgelabel{trans}$} (q0_0);
        \draw[intra] (q0_0) to node [above] {\tiny$\edgelabel{to}$} (q2);
        \draw[intra] (q2)   to node [below]  {\tiny$\edgelabel{trans}$} (q1_1);
        \draw[intra] (q1_1) to node [below]  {\tiny$\edgelabel{to}$} (q3);
        \draw[intra] (q3)   to node [above] {\tiny$\edgelabel{trans}$} (q2_0);
        \draw[intra] (q2_0) to node [above] {\tiny$\edgelabel{to}$} (q2);

  \end{scope}
\end{scope}

\begin{scope}

  \begin{scope}[
      xshift=13cm, 
      yshift=0cm]

    \node[empty] (e0)                              {...};
    \node[evt] (e1) [right of=e0, yshift=-0.5cm] {$e'$};
    \node[evt] (e2) [right of=e1, yshift=-0.5cm] {$e''$};
    \node[evt] (e3) [right of=e2, yshift=-0.5cm] {$e'''$};
    \node[empty] (e4) [right of=e3, yshift=-0.5cm] {...};

    \path[toempty] (e0) edge (e1);
    \path[intra]   (e1) edge node [above] {\tiny$\edgelabel{next}$} (e2);
    \path[intra]   (e2) edge node [above] {\tiny$\edgelabel{next}$} (e3);
    \path[toempty] (e3) edge (e4);

  \end{scope}

  \begin{scope}[
      xshift=13cm, 
      yshift=-4.2cm]

    \node[empty] (c0)                              {...};
    \node[evt] (c1) [right of=c0, yshift=-0.5cm] {$c'$};
    \node[evt] (c2) [right of=c1, yshift=-0.5cm] {$c''$};
    \node[empty] (c3) [right of=c2, yshift=-0.5cm] {...};

    \path[toempty] (c0) edge (c1);
    \path[intra]   (c1) edge node [above] {\tiny$\edgelabel{next}$}  (c2);
    \path[toempty] (c2) edge (c3);

  \end{scope}
\end{scope}

\begin{scope} 
  \begin{scope}[
      xshift=8.3cm, 
      yshift=-0.7cm,
      level 1/.style={sibling distance=1.7cm},
      level 2/.style={sibling distance=1.5cm}, 
      level 3/.style={sibling distance=1.3cm},   
      level distance=1cm]
    
    \node[root] (pt1) {HTTPReq}
      child[intra] { node[term, xshift=0.5cm] {POST} } 
      child[intra] { node[nterm] {res} 
        child[intra] { node[term, yshift=-0.75cm] {/change\_pwd.php}  edge from parent node [left] {\tiny$\edgelabel{child}$}  }  edge from parent node [left] {\tiny$\edgelabel{child}$} 
      }
      child[intra] { node[nterm, xshift=-0.2cm] {hdr.-list}
        child[intra] { node[term, xshift=0.2cm] {SESSION} edge from parent[wave, densely dotted] }
        child[intra] { node[term, xshift=-0.1cm] (sid) {X4a} edge from parent[wave, densely dotted] }
      }
      child[intra] { node[nterm, xshift=0.8cm] {body} 
        child[intra] { node[term] {password} edge from parent[wave, densely dotted] }
        child[intra] { node[term] (pwd1) {pwnd} edge from parent[wave, densely dotted] }
      };
  \end{scope}

  \begin{scope}[
      xshift=9cm, 
      yshift=-4.8cm,
      level 1/.style={sibling distance=1.7cm},
      level 2/.style={sibling distance=1cm}, 
      level 3/.style={sibling distance=1cm},        
      level distance=1cm]
    
    \node[root] (pt2) {SQL-QUERY}
      child[intra] { node[term, xshift=1.2cm] {UPDATE}  edge from parent node [left] {\tiny$\edgelabel{child}$} } 
      child[intra] { node[nterm, xshift=1cm] {trgt-table}
        child[intra] { node[term] {users} }
      }
      child[intra] { node[term, xshift=0.5cm] {SET} }
      child[intra] { node[nterm] {set-cl.-list}
        child[intra] { node[term] {password} edge from parent[wave, densely dotted] }
        child[intra] { node[term] {=} edge from parent[wave, densely dotted] }
        child[intra] { node[term, xshift=-0.1cm] (pwd2) {pwnd} edge from parent[wave, densely dotted] }
      }
      child[intra] { node[term, xshift=-0.4cm] {WHERE} } 
      child[intra] { node[nterm, xshift=-0.8cm] {cond.} 
        child[intra] { node[term, xshift=0.3cm] {sid} edge from parent[wave, densely dotted] }
        child[intra] { node[term, left=-0.2cm]  {=} edge from parent[wave, densely dotted] }
        child[intra] { node[term, xshift=-0.3cm] {X4a} edge from parent[wave, densely dotted] }
      };
  \end{scope}
\end{scope}

\begin{scope}

  \begin{scope}[
      xshift=-4.5cm, 
      yshift=-1cm]
    \node[var] (v1) [xshift=1.5cm] {\begin{tabular}{l} $v_1=\mathtt{X4a}$ \\ \hline  \scriptsize{syn\_type: string}\\  \scriptsize{sem\_type: SU}  \end{tabular}};
    \node[var] (v2) [below=of v1, yshift=0.5cm]  {\begin{tabular}{l} $v_2=\mathtt{pwnd}$ \\ \hline  \scriptsize{syn\_type: string}\\  \scriptsize{sem\_type: UG}  \end{tabular}};

  \end{scope}

  \begin{scope}[
      xshift=-4.3cm, 
      yshift=-5.8cm]
    \node[var] (v3) [below=of v2]     {\begin{tabular}{l} $v_3=\mathtt{X4a}$ \\ \hline  \scriptsize{syn\_type: string}\\  \scriptsize{sem\_type: SU}  \end{tabular}};
    \node[var] (v4) [below=of v3, yshift=0.5cm] {\begin{tabular}{l} $v_4=\mathtt{pwnd}$ \\ \hline  \scriptsize{syn\_type: string}\\  \scriptsize{sem\_type: UG}  \end{tabular}};

  \end{scope}
\end{scope}          

\path[causes] (e1) edge[bend right=5]node[right, color=black] {$\edgelabel{causes}$}  (c1);

\path[parses] (pt1.east) edge[bend left=15] node[above, color=black] {$\edgelabel{parses}$}  (e1.west);
\path[parses] (pt2.east) edge[bend left=15] node[above, color=black] {$\edgelabel{parses}$}  (c1.west);

\path[accepts] (q0_0.north) edge[bend left=30] node[above, color=black] {$\edgelabel{accepts}$} (pt1.west);

\path[propag] (v2.west) edge[bend right=20] node[left, color=black] {$\edgelabel{propag.}$} (v4.west);
\path[propag] (v1.west) edge[bend right=20] node[left, color=black] {$\edgelabel{propag.}$} (v3.west);

\path[ext] (pwd1.south)  edge [bend left=25] node[above, color=black]  {$\edgelabel{source}$}  (v2.east);
\path[ext] (v4.south)    edge [bend right=5] node[above, color=black]  {$\edgelabel{sink}$}   (pwd2.south);

\path[bel] (q2.south) edge [bend left=30] node[below, color=black]  {$\edgelabel{has}$}  (v1.east);
\path[bel] (q2.south) edge [bend left=30] node[below, color=black]  {$\edgelabel{has}$}  (v2.east);
\path[bel] (q2.south) edge [bend left=30] node[below, color=black]  {$\edgelabel{has}$}  (v3.east);
\path[bel] (q2.south) edge [bend left=30] node[below, color=black]  {$\edgelabel{has}$}  (v4.east);

        \end{tikzpicture}
      \end{adjustbox}
\caption{Excerpt of property graphs for a model showing two tiers (logic and data).}
\label{fig:badass_dm}
\end{figure*}
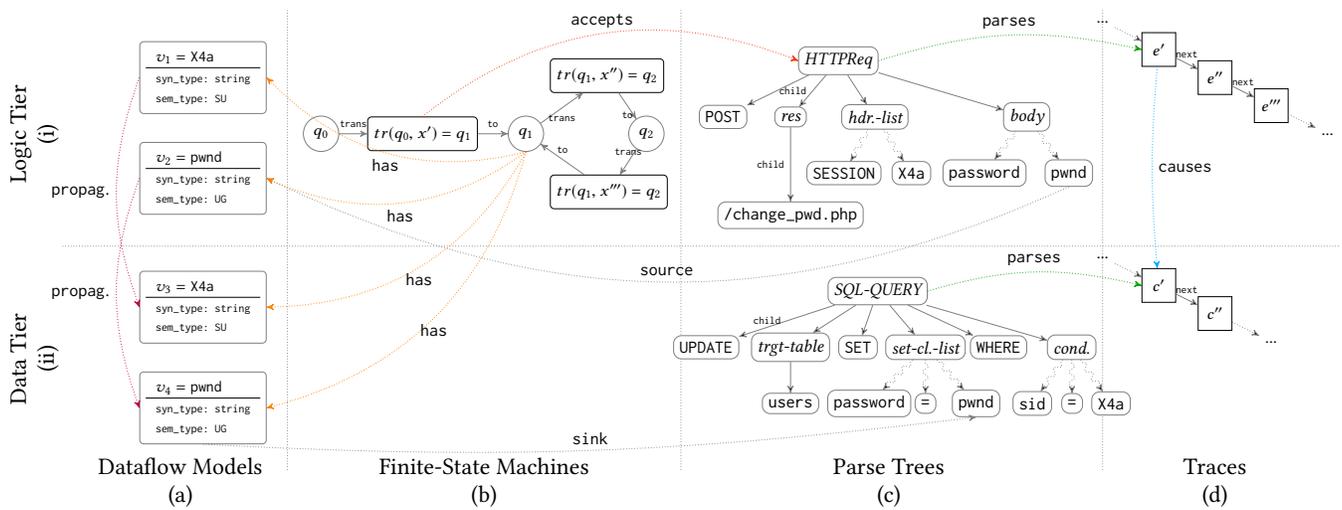

The overall goal of our modeling approach is to create
a representation of a web application that can address challenges C1-3 and
C6. 
Challenge C1 requires obtaining an adequate model that allows determining
when a change of state occurs. We address this challenge by building a
finite-state machine (FSM) from execution traces captured by our probes.
Challenge C2 consists in determining which state transitions are
security-relevant. We observe that security-relevant transitions are likely to
occur less frequently than other transitions. From this observation, we derive
state invariants based on frequency.
Challenge C3 consists in determining the relationship between request
parameters and state transitions. In particular, we are interested in
identifying two types of HTTP parameters: parameters carrying unguessable
tokens and parameters carrying user input. We address this challenge by using
a data flow model (DFM) with types (see~\cite{Wang:2012:SMY:2310656.2310691}).
The DFM represents a state as a set of variables and can capture
the propagation of data items from HTTP requests to the SQL query. Each data
item can have syntactic types, e.g., string, integer, boolean, and semantic
types, e.g., constant, unique, user input. We use types to identify tokens and
user-generated inputs.
Finally, we need a representation for our models that can support (i) the
creation of a model with inference algorithms and (ii) the identification of
security-relevant transitions. To address this challenge, i.e., C6, we
map models into \emph{labeled property graphs} and use \emph{graph traversals}
to query them.

This section details the building blocks of our modeling approach. In
\Cref{sec:deep_model}, we present property graphs, the mapping of models
to graphs, and elementary graph traversals. In
\Cref{sec:construction}, we present the construction of a property graph.

\subsection{Labeled Property Graph}
\label{sec:deep_model}

A labeled property graph is a directed graph in which nodes and edges can have
labels and a set of key-value properties. An example of a labeled property graph
is shown below.
\begin{center}
\begin{adjustbox}{width=0.8\columnwidth}
\begin{tikzpicture}[->, >=stealth', auto, node distance=3cm,]
      \tikzstyle{every state}=[draw=black]

      \node[node] (v1)                {$n_1$:$\nodelabel{L'}$};
      \node[node] (v2)  [right of=v1] {$n_2$:$\nodelabel{L'}$};
      \node[node] (v3)  [right of=v2] {$n_3$:$\nodelabel{L''}$};

      \draw[->,draw=gray] (v1)                      to node [midway] {$e_1$:$\edgelabel{R'}$} (v2);
      \draw[->,draw=gray] ([yshift=0.05cm]v2.east)  to node [midway] {$e_2$:$\edgelabel{R'}$} ([yshift=0.05cm]v3.west);
      \draw[->,draw=gray] ([yshift=-0.05cm]v2.east) to node [below, midway] {$e_3$:$\edgelabel{R''}$} ([yshift=-0.05cm]v3.west);
      \draw[->,draw=gray] (v3)                      to [bend left=20] node [midway] {$e_4$} (v1);

      \node[prop, rectangle split, rectangle split parts=1, draw, text width=1cm] (prop1) [xshift=-0.3cm, yshift=1cm]
        { $k_1$:$v_1$};

      \node[prop, rectangle split, rectangle split parts=1, draw, text width=1cm] (prop2) [xshift=5.7cm, yshift=1cm]
        { $k_2$:$v_2$};

      \draw[-,dashed] (prop1) to (v1);
      \draw[-,dashed] (prop2) to (v3);
\end{tikzpicture}
\end{adjustbox}
\end{center}
\noindent This example shows three nodes. Nodes $n_1$ and $n_3$ have
one property each, i.e., $k_1 = v_1$ for $n_1$ and $k_2 = v_2$ for $n_3$.
Nodes have labels. For example, nodes $n_1$ and $n_2$ are labeled with
$\nodelabel{L'}$ whereas node $n_3$ is labeled with $\nodelabel{L''}$. Edges
are also labeled. The edges $e_1$ and $e_2$ are labeled $\edgelabel{R'}$, and
edge $e_3$ is labeled $\edgelabel{R''}$.

\subsubsection{Mapping Models to Property Graphs}

We now present the mapping of traces, FSM and DFM to a property graph.
\Cref{fig:badass_dm} shows the operation of updating the user password as a
property graph. This example covers the logic and data tiers of a
web application. For the sake of readability, user actions are not shown.

\vspace{4pt}\noindent\textbf{Traces and Parse Trees}---In our approach, traces
and parse trees are important artifacts that are used throughout the analysis.
First, traces and parse trees are the input of the inference algorithms to
generate FSMs and DFMs. Second, traces are used to derive state invariants,
e.g., the number of distinct HTTP requests triggering the same state
transition. Third, parse trees are used for the generation of tests to detect
\acsrf{} vulnerabilities. Accordingly, we decided to include them in the
property graph.

A trace is a sequence of events observed by our sensors, e.g., HTTP messages or
SQL queries. We represent an event with a node of label $\nodelabel{Event}$.
We chain events using edges with label $\edgelabel{next}$. Parse trees
represent the content of a trace event. For example, with reference to
\Cref{fig:badass_dm}.d, the event $e'$ is the following HTTP request:
\begin{lstlisting}[style=http, belowskip=-1 \baselineskip]
POST /change_pwd.php  HTTP/1.1
Host: bank.com
Cookie: SESSION=X4a
Content-Length: 15
Content-Type: application/x-www-form-urlencoded

password=pwnd
\end{lstlisting}
\noindent We parse HTTP requests and store the resulting parse tree in the property
graph. An example of a parse tree for the example is shown in
\Cref{fig:badass_dm}.c.i. For simplicity, \Cref{fig:badass_dm}.c.i does not
show the \texttt{Host}, \texttt{Content-Type}, and \texttt{Content-Length}
HTTP headers. We map parse trees into a property graph as follows. Parse trees
have three labels: $\nodelabel{Root}$, $\nodelabel{NTerm}$, and
$\nodelabel{Term}$. The $\nodelabel{Root}$ node label is used for the root of
a parse tree. The $\nodelabel{NTerm}$ node is used for non-terminal nodes of
the parse tree, whereas $\nodelabel{Term}$ is for the terminal nodes. Nodes
are connected using the $\edgelabel{child}$ edge label.

\vspace{4pt}\noindent\textbf{Finite State Machines}---We use FSMs to represent
program states and transitions between states. Our goal is the identification
of state transitions triggered by an HTTP request. Accordingly, we use HTTP
requests as the symbols accepted by a transition. However, in our model, HTTP
requests are represented as nodes, and property graphs do not support edges
between a node, e.g., an HTTP request, and an edge, e.g., a transition. As a
result, we model a transition between two states as nodes with
three edges. The first edge is directed to the node representing the accepted
HTTP request. The second edge is from the initial state of the transition to
the transition node. The third edge is directed to the new state. The mapping
of FSM elements to nodes, edges, and labels is shown in \Cref{tab:nodes}.

\vspace{4pt}\noindent\textbf{Dataflow Information and Types}---To determine the
relationship between request parameters and state changing operations, we use
dataflow models (DFMs) with types as presented by Wang et
al.~\cite{Wang:2012:SMY:2310656.2310691}. The data flow model was originally
designed to enrich HTTP request parameters with abstract types such as
syntactic and semantic tables. Consider an HTTP request with a parameter
\texttt{password=pwnd} with the value \texttt{pwnd} provided by the
user. The DFM associates the parameter \texttt{password} with a syntactic
label, e.g., \texttt{string}, and semantic labels, for example, user-generated
(\texttt{UG}). In our graph, we represent a DFM as a set of variables. A
variable is a node graph with a name (e.g., parameter name), a value (e.g.,
parameter value), and a type (e.g., semantic and syntactic type). Variables
can carry the same data item. In these cases, we say that there is a
propagation of data values. The rules that determine whether a propagation
exists are presented in \Cref{sec:construction}.

An example of a DFM is shown in \Cref{fig:badass_dm}.a. This DFM comprises
four variables, two for HTTP request parameters, i.e., session cookie and
password parameter, and two for the SQL WHERE and SET clauses. Each variable
has a type. For example, variable $v_1$ has semantic type \texttt{SU}, which
means that the value is different for each user session, whereas varuable
$v_2$ has type \texttt{UG}. We represent the propagation of data items with a
source, a propagation chain and a sink. For this, we use three types of edges,
$\edgelabel{source}$, $\edgelabel{propag.}$, and $\edgelabel{sink}$.
\Cref{fig:badass_dm} shows the complete propagation chain for the
$\texttt{pwnd}$ data item. Finally, DFM variables are linked to FSM states
with $\edgelabel{has}$ edges. This link determines the relationship between
request parameters and state-changing operations.

\begin{table*}
\begin{minipage}{.5\linewidth}
  \footnotesize
    \centering
    \begin{tabular}{l l L{1.6cm}}
    \toprule
    Component   & Node label(s)   & Relationship(s) \\
    \midrule
    FSM         & $\nodelabel{State}$, $\nodelabel{StateTrans}$                & $q \xrightarrow{\text{trans}} t$, $t \xrightarrow{\text{to}} q$, $t \xrightarrow{\text{accept}} q$\\
    DFM         & $\nodelabel{Variable}$                                       & $v' \xrightarrow{\text{propagat}} v''$\\
    Trace       & $\nodelabel{Event}$                                          & $e' \xrightarrow{\text{next}} e''$\\
    Parse tree  & $\nodelabel{Root}$, $\nodelabel{NTerm}$, $\nodelabel{Term}$  & $n \xrightarrow{\text{child}} m$\\
    \bottomrule
    \end{tabular}
    \caption{List of nodes and edges for our models.}
    \label{tab:nodes}
\end{minipage}
\begin{minipage}{.5\linewidth}
  \footnotesize
    \centering
    \begin{tabular}{l L{5.5cm}}
    \toprule
    Name                    & Mapping into a Property Graph\\
    \midrule
    Data Flow Inform.        & $v:\nodelabel{State} \xrightarrow{\text{has}} q:\nodelabel{Variable}$\\
    Data Propagation        & $v_1:\nodelabel{Variable} \xrightarrow{\text{propag.}} v_2:\nodelabel{Variable}$ or $t:\nodelabel{Term}$\\
    Abstractions            & $apt:\nodelabel{Root} \xrightarrow{\text{abstracts}} pt:\nodelabel{Root}$, $ae:\nodelabel{Event} \xrightarrow{\text{abstracts}} e:\nodelabel{Event}$\\
    Event Causality          & $e_1:\nodelabel{Event} \xrightarrow{\text{causes}} e_2:\nodelabel{Event}$\\
    Accepted Inputs         & $st:\nodelabel{StateTrans} \xrightarrow{\text{accepts}} pt:\nodelabel{Root}$\\
    \bottomrule
    \end{tabular}
    \caption{List of relationships between models.}
    \label{tab:relationships}
\end{minipage}
\end{table*}

\subsubsection{Relationships}

The elements of our graph have relationships. Consider, for example, a parse
tree that represents the HTTP request causing a state transition. Our
framework defines a set of relationships between these elements. We now
briefly present these relationships. The mapping of these relationships into a
property graph is shown in \Cref{tab:relationships}.\gp{Link the text to the
example}

\vspace{4pt}\noindent\textbf{Dataflow Information}---This relationship connects a
DFM to a FSM, or a DFM to a parse tree. In the first case, the variable can be
used to determine the state of a FSM. We model this relationship with an edge
$\label{has}$ from a state to a variable. In the second case, a variable
carries values from a source, e.g., HTTP parameters, or values used to create
a query.

\vspace{4pt}\noindent\textbf{Data Propagation}---This relationship captures the
propagation of data items during the execution of a program. In our model,
this relationship is between two DFMs and represents the propagation of data
items across the tiers of a web application. For example, consider a data
value that is first provided with a user action; then the value is included in
an HTTP request; and, finally, it is inserted in a SQL query to be stored in
the database. 

\vspace{4pt}\noindent\textbf{Abstractions}---Abstractions represent the
link between an abstract element and its concrete counterpart. Abstractions
are an expedient to reduce the complexity of a problem or to focus the
analysis on relevant parts. For example, abstractions remove variable parts
such as data values from SQL queries. The resulting abstract SQL query is then
compared with other abstract queries to group them. This expedient is used by
our model inference algorithms and we present abstractions in
\Cref{sec:construction}.

\vspace{4pt}\noindent\textbf{Event Causality}---This relationship can occur, for
example, between a user click on a link and the resulting HTTP request. Our
sensors can establish this type of relationship.

\vspace{4pt}\noindent\textbf{Accepted Inputs}---This relationship captures the
connection between HTTP requests and state transitions. Iff HTTP requests 
cause a transition, we say that the FSM accepts the HTTP request.

\subsubsection{Graph Traversals}

Graph traversals are the means to retrieve information from property graphs.
They allow querying a graph based on nodes, edges, and properties. \tool{}
uses traversals written in the Cypher query language~\cite{cypher}, a graph
query language supported by popular graph databases such as Neo4j. The Cypher
language follows a declarative approach in which each query describes
\emph{what} we want to retrieve and not \emph{how}. The \emph{what} is
specified with \emph{graph patterns}, a description of a subgraph using nodes,
edges, labels, and properties. \tool{} uses graph queries for the creation of
FSM and DFS (See \Cref{sec:construction}) and to generate tests for the
detection of \acsrf{} (See \Cref{sec:mining}).

For the sake of readability, we do not present the Cypher syntax but a simplified
notation that retains the declarative approach. We use sets of nodes and edges
to represent Cypher queries. For example, a query $Q$ can be defined as all
nodes $n$ in the property graph for which a given predicate $p$ is true, i.e.,
$Q=\{n: p(n) \}$. In our notation, the predicate $p$ is the graph pattern. We
use parametric logic predicates for graph patterns. In the following, we
present elementary graph patterns that allow establishing a basic language to
operate with the property graph.

We start with an example to show elementary queries to retrieve nodes and
edges via labels. These queries are generic and are not tied to our framework.

\begin{example}[Elementary Queries]
\label[example]{ex:query1}
  To create queries, we first define the graph pattern. Then, we use the
  predicate to define a set. The first elementary pattern is true iff a node
  has a given label $\nodelabel{L}$:
  \[
  \begin{split}
  \QLabel_{\nodelabel{L}}(n) \eqdef & ``n:\nodelabel{L} "
  \end{split}
  \]
  \noindent The second example pattern is true iff a graph edge has a
  given label $\mathcal{R}$:
  \[
  \begin{split}
  \QLabel_{\mathcal{R}}(n, m) \eqdef & ``e=(n, m) \wedge e:\mathcal{R} "
  \end{split}
  \]
  These predicates can be used to define queries. For example, to find all
  nodes with label $\nodelabel{L}$ we can write the following query:
  \[ \Q{Q_{label}} = \{ n: \QLabel_{\nodelabel{L}}(n) \} \]
  As graph patterns may have more than one parameter, we can use quantifiers
  (i.e., $\forall$ or $\exists$) to broaden or limit the scope of a query. For
  example, consider the query to retrieve \emph{all} nodes with an outgoing
  edge   $\edgelabel{R}$, we can use the following query:
  \[ \Q{Q_{out}} = \{ n:\forall m, \QLabel_{R}(n, m) \} \]

\end{example}

From these elementary patterns and queries, we create a basic query language
that can express elements of our models.

\begin{example}[Queries for Models]
\label[example]{ex:query2}

  Consider the example of retrieving the states of a FSM. First, we define a
  predicate for the pattern, called $\QState(q)$, that is defined as
  $\QLabel_{\edgelabel{State}}(n)$. Then, we use this pattern in a query that
  searches for all states $q$:
  \[
  \begin{split}
  \Q{Q_{States}} \eqdef  & \{ q: \QState(q) \}
  \end{split}
  \]
  \noindent We create similar patterns for relationships. For example, with
  reference to \Cref{fig:badass_dm}, consider the graph pattern between the
  state $q_0$ and $q_1$. We can call this pattern $\QTrans(q_0, t, q_1)$ and
  we define it as $\QLabel_{\edgelabel{trans}}(q_0, t) \wedge
  \QLabel_{\edgelabel{to}}(t, q_1)$.

  In a similar way, we create patterns for all nodes and edges in
  \Cref{tab:nodes} and in \Cref{tab:relationships}. We also create patterns
  using properties. For example, $\QHTTPReq(pt)$ is a pattern for a
  $\nodelabel{Root}$ node $pt$ whose property
  $\propname{t}=\propvalue{HttpReq}$. This gives us a basic language to
  operate with our models.

\end{example}

The notation of these two examples adheres to the declarative approach followed
by Cypher. The actual search of all nodes matching the predicates used in the
set definition is performed by the query processor. The query processor is a
graph database component that transforms declarative queries into a sequence
of operations to traverse the graph and search for all matching nodes.

\subsection{Model Construction}
\label{sec:construction}

After having presented the building blocks of our modeling approach, we
present the construction of our model. The first step of the construction
consists in importing traces and parse trees in the property graph. Then, we
use inference algorithms to create FSMs and DFMs. 

\subsubsection{Importing Traces and Parse Trees} We import
traces and parse trees in the following order:

\vspace{4pt}\noindent\textbf{User Actions}---We first import user actions
traces. For each element of the trace, we create a node $\nodelabel{Event}$.
If two events are consecutive in a trace, then we place an edge
$\edgelabel{next}$ between the two nodes. Then, we parse the user action into
the three main elements: the type of action (e.g., mouse click or key stroke),
the UI element on which the action is performed (e.g., HTML element), and, if
present, the user input (e.g., username). Then, we connect the root node of
the parse tree to the trace node with a $\edgelabel{parses}$ edge. To
distinguish user action events from other events (i.e., HTTP messages), we add
a node property $\propname{t}$ to $\propvalue{UA}$ which stands for user
action. Finally, we add a node property for the user performing these actions.
For example, if the user actions are performed by an administrator, we add the
property $\propname{user}=\propvalue{admin}$.

\vspace{4pt}\noindent\textbf{HTTP Messages}---First, we import a
trace as seen for user actions. Second, for each HTTP message, we create parse
trees for HTTP requests, responses, URLs, cookies, HTTP POST data, and JSON
objects. We link the root with the event with a
$\edgelabel{parses}$ relationship. Then, we link the HTTP messages to network
events with $\edgelabel{parses}$ edges, and $\edgelabel{causes}$ edges between
user actions and HTTP request events. The property $\propname{t}$ is set to
$\propvalue{HTTPReq}$. Finally, as described in \Cref{sec:detection}, \tool{}
reproduces user actions twice, thus generating two HTTP message traces, i.e.,
sessions, which can be different due to newly generated cookies or
anti-CSRF tokens. When importing traces, we add the trace session number as a
node property.

\vspace{4pt}\noindent\textbf{Database Queries}---We parse the call trees
to extract calls to database APIs and retrieve SQL queries. We add a
$\edgelabel{parses}$ relationship between the parse trees and the trace event.
Then, we add causality edges between HTTP request events and the resulting
query events. Similarly as for HTTP messages, we add the trace session number
as a node property. Finally, the property $\propname{t}$ is set to
$\propvalue{SQL}$.

\subsubsection{Finite-State Machines} After importing traces
and creating parse trees, we construct the FSM.

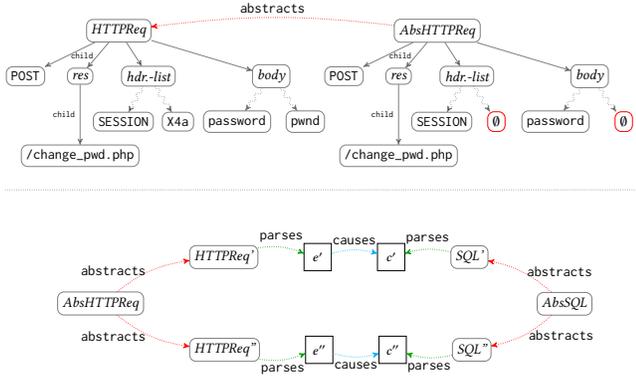
\begin{figure}[t!]
\centering
\begin{adjustbox}{width=1\columnwidth}
  \begin{tikzpicture}

\draw[color=gray, densely dotted] (-2.5, -3.5) -- (11.5,-3.5);

  \begin{scope}[
      xshift=0cm, 
      yshift=0cm,
      level 1/.style={sibling distance=1.7cm},
      level 2/.style={sibling distance=1.5cm}, 
      level 3/.style={sibling distance=1.3cm},   
      level distance=1cm]
    
    \node[root] (pt1) {HTTPReq}
      child[intra] { node[term, xshift=0.5cm] {POST} } 
      child[intra] { node[nterm] {res} 
        child[intra] { node[term, yshift=-0.75cm] {/change\_pwd.php}  edge from parent node [left] {\tiny$\edgelabel{child}$}  }  edge from parent node [left] {\tiny$\edgelabel{child}$} 
      }
      child[intra] { node[nterm, xshift=-0.2cm] {hdr.-list}
        child[intra] { node[term, xshift=0.2cm] {SESSION} edge from parent[wave, densely dotted] }
        child[intra] { node[term, xshift=-0.1cm] (sid) {X4a} edge from parent[wave, densely dotted] }
      }
      child[intra] { node[nterm, xshift=0.8cm] {body} 
        child[intra] { node[term] {password} edge from parent[wave, densely dotted] }
        child[intra] { node[term] (pwd1) {pwnd} edge from parent[wave, densely dotted] }
      };
  \end{scope}

  \begin{scope}[
    xshift=7cm, 
    yshift=0cm,
    level 1/.style={sibling distance=1.7cm},
    level 2/.style={sibling distance=1.5cm}, 
    level 3/.style={sibling distance=1.3cm},   
    level distance=1cm]
  
  \node[root] (apt1) {AbsHTTPReq}
    child[intra] { node[term, xshift=0.5cm] {POST} } 
    child[intra] { node[nterm] {res} 
      child[intra] { node[term, yshift=-0.75cm] {/change\_pwd.php}  edge from parent node [left] {\tiny$\edgelabel{child}$}  }  edge from parent node [left] {\tiny$\edgelabel{child}$} 
    }
    child[intra] { node[nterm, xshift=-0.2cm] {hdr.-list}
      child[intra] { node[term, xshift=0.2cm] {SESSION} edge from parent[wave, densely dotted] }
      child[intra] { node[term, xshift=-0.1cm, draw=red] (sid) {$\emptyset$} edge from parent[wave, densely dotted] }
    }
    child[intra] { node[nterm, xshift=0.8cm] {body} 
      child[intra] { node[term] {password} edge from parent[wave, densely dotted] }
      child[intra] { node[term, draw=red] (pwd1) {$\emptyset$} edge from parent[wave, densely dotted] }
    };
\end{scope}

\begin{scope}[
   xshift=-0.4cm, 
    yshift=-6cm,
    node distance=3cm]
      \node[root] (apt3) {AbsHTTPReq};
      \node[root] (pt31) [above right=0.5cm and 1cm of apt3] {HTTPReq'};
      \node[root] (pt32) [below right=0.5cm and 1cm of apt3] {HTTPReq''};
      \node[evt]  (e31)  [right=1cm of pt31] {$e'$};
      \node[evt]  (e32)  [right=1cm of pt32] {$e''$};
      \node[evt]  (d31)  [right=1cm of e31]  {$c'$};
      \node[evt]  (d32)  [right=1cm of e32] {$c''$};
      \node[root] (pt41) [right=1cm of d31] {SQL'};
      \node[root] (pt42) [right=1cm of d32] {SQL''};
      \node[root] (apt4) [above right=0.5cm and 1cm of pt42] {AbsSQL};
\end{scope}

\path[causes] (e31) edge[bend left=15] node[above, color=black] {$\edgelabel{causes}$}  (d31);
\path[causes] (e32) edge[bend right=15] node[below, color=black] {$\edgelabel{causes}$}  (d32);

\path[abstracts] (apt3) edge[bend left=15] node[left, color=black] {$\edgelabel{abstracts}$}  (pt31);
\path[abstracts] (apt3) edge[bend right=15] node[left, color=black] {$\edgelabel{abstracts}$}  (pt32);
\path[abstracts] (apt4) edge[bend right=15] node[right, color=black] {$\edgelabel{abstracts}$}  (pt41);
\path[abstracts] (apt4) edge[bend left=15] node[right, color=black] {$\edgelabel{abstracts}$}  (pt42);

\path[abstracts] (apt1) edge[bend right=7] node[above, color=black] {$\edgelabel{abstracts}$}  (pt1);

\path[parses] (pt31) edge[bend left=15] node[above, color=black] {$\edgelabel{parses}$}  (e31);
\path[parses] (pt32) edge[bend right=15] node[below, color=black] {$\edgelabel{parses}$}  (e32);
\path[parses] (pt41) edge[bend right=15] node[above, color=black] {$\edgelabel{parses}$}  (d31);
\path[parses] (pt42) edge[bend left=15] node[below, color=black] {$\edgelabel{parses}$}  (d32);

  \end{tikzpicture}
\end{adjustbox}
\caption{On top: abstract relationships between a parse tree and
an abstract one. Below: visualization of the graph pattern to identify transitions.}
\label{fig:abs_tree}
\end{figure}

\vspace{4pt}\noindent\textbf{Abstract Parse Trees}---The rule to build a FSM is
the following: A state transition occurs when similar HTTP requests cause
similar SQL queries. Similarity between HTTP requests and queries is achieved
by the means of \emph{abstract parse trees}, i.e., parse trees that omit a few
selected terminal nodes. For HTTP requests, we neglect URL
parameter values and POST data values. For SQL queries, we neglect terminal
nodes at the right-hand side of SQL comparison operations.
\Cref{fig:abs_tree} shows the parse
tree of an HTTP request to update a user password and an abstract parse tree in
which terminal nodes were neglected.  Abstract parse trees are
unique. If two parse trees result in the same abstract tree, we place two
edges $\edgelabel{abstracts}$ from the abstract parse tree to the two parse
trees.

\vspace{4pt}\noindent\textbf{Clustering}---After the creation of abstract
parse trees, we extract HTTP requests triggering the same transition from the
graph . \Cref{fig:abs_tree} exemplifies this situation, showing the roots of
parse trees and trace events. Two requests, e.g., the roots $HTTPReq'$ and
$HTTPReq''$, trigger the same transition if (i) the HTTP requests have the
same abstract parse tree, i.e., with root $AbsHTTPReq$, (ii) the HTTP requests
cause SQL queries, i.e., parse tree roots $SQL'$ and $SQL''$, via a causality
edge, and (iii) the SQL queries have the same abstract parse tree, i.e.,
$AbsSQL$. HTTP requests matching this description can be found with this
query:
\[
\begin{split}
  \Q{Q_{Aux}} \eqdef & \{ (abs_h', h', abs_{sql}', sql'): \exists e', c', \QAbs(abs_h', h') \wedge \\
                       & \QParses(h', e') \wedge \QCaused(e', c') \wedge\\
                       & \QParses(sql', c') \wedge \QAbs(abs_{sql}', sql') \}
  \end{split}
\]

\vspace{2pt}\noindent This query returns a set of 4-tuples. For example, with reference to
\Cref{fig:abs_tree}, this query returns two 4-tuples: the first with
$AbsHTTPReq'$, $HTTPReq'$, $AbsSQL'$, and $SQL'$, and the second with
$AbsHTTPReq''$, $HTTPReq''$, $AbsSQL''$, $SQL''$. If we group these tuples by
abstract HTTP request and abstract SQL query, the resulting groups represent
transitions satisfying our rule. The HTTP requests in each group are the
symbols causing the state transition.

\vspace{4pt}\noindent\textbf{FSM}---To create a FSM, we create one state node
for each edge $\edgelabel{next}$, and a transition for each HTTP request.
Then, we minimize the FSM using the clustering
algorithm~\cite{Hopcroft_automata}.

\subsubsection{Dataflow Model and Information} Finally, we construct
the data flow model with types.

\vspace{4pt}\noindent\textbf{Variables}---Variables are derived from terminal
nodes in parse trees. The terminal nodes are the same ones neglected in
abstract parse trees. The value of the variable is the symbol of the
terminal node, whereas the variable name is the path of the terminal node from
the root. Then, we link variables to states with an edge $\edgelabel{has}$.

\begin{figure}[t]
\centering
\begin{adjustbox}{width=1\columnwidth}
  \trimbox{0cm 0cm 0cm 0cm}{
      \begin{tikzpicture}[->, >=stealth', auto, node distance=2cm]
        \tikzstyle{every state}=[draw=black]
        \node[evt, draw=none] (dots0)                  {};
        \node[evt]            (e2) [right of=dots0]    {$e_1$};
        \node[evt]            (e3) [right of=e2]       {$e_2$};
        \node[evt, draw=none] (dots1) [right of=e3]    {};
        \path[->,dashed,draw=gray] (dots0) edge(e2)
                                   (e3)    edge (dots1);
        
        \path[intra] 
              (e2)    edge node {$\edgelabel{next}$} (e3);
              
        \node[evt, draw=none] (dots2) [below of=dots0,  yshift=-0.3cm] {};
        \node[evt]            (ep1)   [below of=e3,     yshift=-0.3cm]    {$e'_2$};
        \node[evt, draw=none] (dots3) [below of=dots1,  yshift=-0.3cm] {};

        \path[->,dashed,draw=gray] (dots2) edge  (ep1)
                                   (ep1)   edge  (dots3);  
        
        \node[evt, draw=none] (dots4) [below of=dots2,  yshift=-0.3cm] {};
        \node[evt]            (epp1)  [below of=ep1,    yshift=-0.3cm]   {$e''_2$};
        \node[evt, draw=none] (dots5) [below of=dots3,  yshift=-0.3cm] {};

        \path[->,dashed,draw=gray] (dots4) edge (epp1)
                                   (epp1)   edge  (dots5);  

        \path[causes] (e3)   edge[bend left=10] node {$\edgelabel{causes}$} (ep1);   
        \path[causes] (ep1)  edge[bend left=10] node {$\edgelabel{causes}$} (epp1);   

        \node[root] (ua1) [below of=e2,  yshift=1.0cm, xshift=-1.4cm] {UserAction$_1$};
        \node[term] (v1)  [below of=ua1, yshift=1.4cm, xshift=-1.4cm] {pwnd};
        \path[parses] (ua1)  edge[bend left=10] node[below right] {$\edgelabel{parses}$} (e2); 
        \path[wavenodec, densely dotted] (v1)   edge[decorate] node[below right] {$\edgelabel{child}$} (ua1); 

        \node[root] (http1) [below of=e3,   yshift=1.0cm, xshift=-1.4cm]   {UserAction$_2$};
        \path[parses] (http1)  edge[bend left=10] node[below right] {$\edgelabel{parses}$} (e3); 

        \node[root] (http2) [below of=ep1,   yshift=1.0cm, xshift=-1.4cm]  {HTTPReq};
        \node[term] (v2)    [below of=http2, yshift=1.4cm, xshift=-1.4cm]  {pwnd};
        \path[parses] (http2)  edge[bend left=10] node[below right] {$\edgelabel{parses}$} (ep1); 
        \path[wavenodec, densely dotted] (v2)   edge[decorate] node[below right] {$\edgelabel{child}$} (http2); 
        
        \node[root] (sql1) [below of=epp1,   yshift=1.0cm, xshift=-1.4cm]  {SQL};
        \node[term] (v3)   [below of=sql1,   yshift=1.4cm, xshift=-1.4cm]  {pwnd};
        \path[parses] (sql1)  edge[bend left=10] node[below right] {$\edgelabel{parses}$} (epp1); 
        \path[wavenodec, densely dotted] (v3)   edge[decorate] node[below right] {$\edgelabel{child}$} (sql1); 

        \path[-,draw=red, thick] (v1.west) edge[bend right=45] node[left, yshift=-0.4cm, text=red] {Case 2} (v2.west);

        \path[-,draw=blue, thick] (v2.west) edge[bend right=25] node[left, yshift=-0.5cm, text=blue] {Case 1} (v3.west);

        \node[draw=none] (ua)[above of=dots0, yshift=-2.0cm, xshift=-3.5cm]   {{(a) User action trace}};
        \node[draw=none] (ua)[above of=dots2, yshift=-2.0cm, xshift=-3.3cm]  {{(b) HTTP message trace}};
        \node[draw=none] (ua)[above of=dots4, yshift=-2.0cm, xshift=-3.45cm]  {{(c) DB queries trace}};

      \end{tikzpicture}
    }
\end{adjustbox}
\caption{Example of propagation along causality edges (Case 1) and backward propagation chain (Case 1).}
\label{fig:backw_prop}
\end{figure}
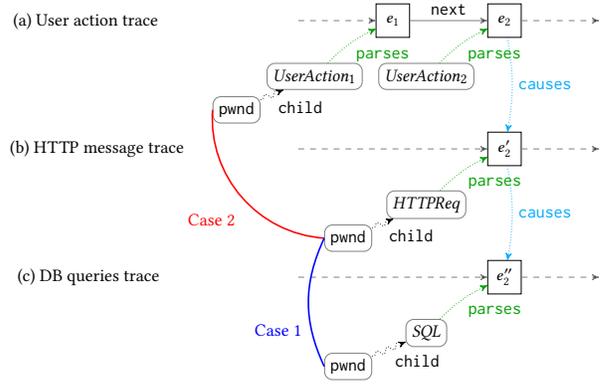

\vspace{4pt}\noindent\textbf{Data Propagation}---After the creation of variable
nodes, we reconstruct the propagation of data values traversing application
tiers. Consider the example in \Cref{fig:backw_prop} which models a user
changing her password. The user types a new password \texttt{pwnd} via a user
action, i.e., $e_1$. This user action is parsed by the parse tree with root
$UserAction_1$. Then, the user submits the password ($e_2$) which is received
by the server ($e'_2$) in an HTTP request with root $HTTPReq$. Finally, the
server uses the password in a query ($e''_2$) with root $SQL$. In this
example, we can distinguish two cases of data propagation. In the first case,
the data item \texttt{pwnd} propagates along causality edges, i.e., from
$e'_2$ to $e''_2$. In these cases, we create a query to retrieve terminal
nodes of HTTP and SQL trees that are reachable via causality edges as shown in
\Cref{fig:backw_prop}. The variables associated to these terminal nodes are
then linked via a $\edgelabel{propag.}$ edge. In the second case, the data
items propagates from $e_1$ to $e_2'$ using first an edge $\edgelabel{next}$,
and then a causality edge. We create a query to retrieve the terminal nodes
from user actions to HTTP requests using the query pattern in
\Cref{fig:backw_prop}, and then we place $\edgelabel{propag.}$ edges between the
variables.

\vspace{4pt}\noindent\textbf{Type Inference}---We use types to distinguish
security-relevant data values (e.g., anti-CSRF tokens) from uninteresting
ones (e.g., constants). Starting from a state transition, we select all
variables of a state and group by variable name. Each group is passed to a
type inference algorithm which returns the types matching each group. The type
inference extracts both syntactical types, e.g., integer, decimal, and boolean,
and semantic ones, e.g., session unique (\texttt{SU}), user unique
(\texttt{UU}) and constant (\texttt{CO}). The rules to infer a semantic type
are the following. If all values are the same, then the type is \texttt{CO}.
If the data values are the same within a trace session but different between
sessions, then the type is \texttt{SU}. If the data values are the same within
the traces of a user, but different between users, then the type is user
unique, i.e., \texttt{UU}. The user-generated (\texttt{UG}) semantic type is
added when there is a propagation chain that starts from a user action. For
example, the chain for \texttt{pwnd} is of type \texttt{UG}.

\section{Model Mining and Test Execution}
\label{sec:mining}

We now present the test generation via model mining (\Cref{sec:tg}) and the
process of test execution and evaluation (\Cref{sec:texec}).

\subsection{Test Generation}
\label{sec:tg}

A test of our approach is a state-changing HTTP request and, optionally, an
HTTP request parameter carrying an anti-CSRF token. First, we query our model
to retrieve all relevant state-changing HTTP requests. Second, for each HTTP
request, we mine our model to retrieve HTTP parameter names that carry an
anti-CSRF token. As a final step, we query our model to extract the
\emph{oracle}. The oracle represents expected behavior that we need to observe
during a test to decide whether a relevant state transition occurred.

We begin with a query to detect HTTP requests that trigger security-relevant
state transitions. Then, we present the query to identify parameters\cro{what
tokens?}. Finally, we present a traversal to extract the test oracle.

\subsubsection{State Transitions} State-changing HTTP requests can
be retrieved by starting from all state transition nodes, and then by
traversing the $\edgelabel{accepts}$ to reach an HTTP request. If such an edge
exists, then the HTTP request is causing a change of state. We can express
this graph traversal as follows. The graph pattern representing connections
between an HTTP request parse tree $pt$, and a state transition node $t$, is
the following:
\[
\begin{split}
\QSC(pt, q', tr, q'') \eqdef & \QTrans(q', tr, q'') \wedge \QFSMAccepts(tr, pt) \wedge\\ 
                                    & \QHTTPReq(pt)                 \\
\end{split}
\]
\noindent where $q'$ and $q''$ are the two states involved in the state
transition $tr$ and $pt$ is an HTTP request. Then, we use the predicate in a
query:
\[
\begin{split}
\Q{Q_{SC}} \eqdef & \{ pt : \forall q', q'', tr, \QSC(pt, q', tr, q'')\}
\end{split}
\]
\noindent This set contains all parse tree roots $pt$ that can trigger
\emph{any} transition of state. \cro{Do you really mean pt, or abspt here?}

\subsubsection{Relevant State Transitions} $\Q{Q_{SC}}$
contains all HTTP requests that cause a change of state. However, not all
changes of state are relevant. For example, requests may result in database
operations to log user activities, which is not a security-critical action. To
identify such non-critical state changes, we hypothesize that irrelevant
queries are likely to occur multiple times within a trace. The occurrence of
queries can be determined via abstract parse trees for queries. As a result of
the FSM construction, all SQL parse trees reachable via
$\edgelabel{abstracts}$ from the same abstract SQL query are \emph{similar}
queries. The number of outgoing $\edgelabel{abstracts}$ edges is the number of
occurrences of similar queries.

Starting from this observation, we refine $\Q{Q_{SC}}$ to take into account
abstract parse trees of SQL queries and their outgoing $\edgelabel{abstracts}$
edges. The refinement extends $\Q{Q_{SC}}$ by traversing (i) an edge
$\edgelabel{parses}$ from the HTTP request to the HTTP message event, (ii) a
causality edge from HTTP message to the data layer event, (iii) a
$\edgelabel{parses}$ edge from the data event to the SQL query, and (iv) the
SQL query to the abstract SQL query. This query returns a list of pairs of the
root of an HTTP request and the root of an abstract SQL query. From this list,
we remove all pairs whose abstract SQL query has a number of outgoing edges
greater than 1. The HTTP requests of the remaining pairs are called
\emph{relevant} state change transitions. We show the accuracy of this
heuristic in \Cref{sec:evaluation}.

\subsubsection{Security Tokens} After having identified
relevant state-changing requests, we search for parameters carrying anti-CSRF
tokens. Anti-CSRF tokens can be transported as URL parameters, POST
parameters, or in custom HTTP headers. During the construction of the DFM, we
created variables with semantic types. For example, variables labeled as
\texttt{SU} or \texttt{UU} carry a value that changes across sessions. As
anti-CSRF tokens are required to be unpredictable for the attacker, these
variables can carry these tokens. For each state-changing HTTP request, we
select all variables with type \texttt{SU} or \texttt{UU}. Given the root of
the parse tree of an HTTP request, we traverse the $\edgelabel{accepts}$ to
reach the transition node. From the transition node, we traverse the
$\edgelabel{to}$, thus reaching the new state. Then, we retrieve all variables
with $\propname{sem\_type} \in \{ \propvalue{UU}, \propvalue{SU} \}$. The
output of these queries is a list of pairs of a state-changing HTTP request
and a variable name carrying a potential anti-CSRF token.

\subsubsection{Oracle} The HTTP request and, optionally, the parameter
carrying an anti-CSRF token are used to generate a test against the web
application. At the end of a test, we need a way to establish whether a
security-relevant state transition occurred. As discussed, a state transition
is relevant if it executes a non-reoccurring SQL query. Accordingly, for each
HTTP request that we intend to test, we retrieve the abstract parse tree
roots of SQL queries with an out-degree equal to one. The traversal to reach
abstract SQL queries is shown in \Cref{fig:abs_tree}. These abstract SQL
queries are the oracle for the HTTP request.

\subsection{Security Tests}
\label{sec:texec}

We now have pairs of parse trees of state-changing HTTP requests and
parameters. The goal of our security tests is to verify the replay-ability of the
requests and check whether they cause SQL queries that are similar to ones in
the oracle.

We test web applications as follows. If the HTTP request has an anti-CSRF
parameter, we generate an HTTP request by omitting the parameter. If the HTTP
request does not have an anti-CSRF parameter, we generate an HTTP request from
scratch. In both cases, we update the request's session cookie by replaying
the user login user actions\footnote{User actions traces are factored in two
parts: actions for the user login and actions for the web application
operation. Existing tools to capture user actions, e.g., Selenese
IDE~\cite{selenium}, support trace factoring. Factoring can be done during the
capture or after the generation by searching for user credentials in the
trace. We detail the creation of factored user actions traces in
\Cref{sec:evaluation}.}. During the test execution, we retrieve the resulting
server-side call graph trace to extract SQL queries. Then, we compare SQL
queries with our oracle. The comparison can result in one of the following
cases. If one of the observed queries matches a relevant query of our model,
then our test managed to reproduce the same change of state. In this case, we
mark the test as \emph{successful}. If all queries either match a repeated
query or are not in our model, then we conclude that we cannot reproduce the
same state-changing operation, and mark the test as \emph{failed}.

\section{Evaluation}
\label{sec:evaluation}

We now present the evaluation of \tool{} against popular web applications.

\subsection{Testbed}
\label{sec:testbed}

\begin{table}[t]
  \footnotesize
  \centering
  \begin{tabular}{l l r r}
    \toprule
    Category   & Web Application & \multicolumn{1}{c}{Version} & \multicolumn{1}{c}{LoC}\\
    \cmidrule(lr){1-4}
    Accounting & Invoice Ninja (IN)              & 2.5.2     & 1,576,957 \\
               & Simple Invoices (SI)            & 2013.1b.8 &   601,532 \\

    \cmidrule(lr){1-4}
    eCommerce  & AbanteCart                      & 1.2.4   &   151,807 \\
               & OpenCart                        & 2.1.0   &   153,863 \\
               & OXID eShop                      & 4.9.8   &   370,723 \\
               & PrestaShop                      & 1.6.1.2 &   420,626 \\
     
    \cmidrule(lr){1-4}
    Forum      & MyBB                            & 1.8.8   &   150,622 \\
               & Simple Machines Forum (SMF)     & 2.0.12  &   153,072 \\
    
    \cmidrule(lr){1-4}
    eMail      & Horde Groupware Webmail (Horde) & 5.12.14 &   178,880 \\
               & Mautic                          & 1.4.1   & 2,190,920 \\
    \bottomrule

  \end{tabular}
  \caption{Web applications for the evaluation.} 
    \label{tab:tbedlist}
\end{table}

We assessed \tool{} against ten web applications retrieved from the Bitnami
catalog~\cite{bitnami}. Bitnami is a provider of packaged, ready-to-deploy
applications that are typically created upon a customer request. Based on this
model, we consider the Bitnami catalog to contain popular web applications.

We selected web applications from four categories, i.e., accounting,
eCommerce, email, and forum, in order of appearance. We collected initially 20
applications. Then, during the instrumentation and trace generation, we
decided to discard 10 of them: Four used an unsupported runtime environment
(i.e., Java or Python), two required paying fees, three of them suffered from
a bug in Xdebug (an important component for our approach), and
one required a publicly available email server. The list of selected web
applications is shown in \Cref{tab:tbedlist}.

\subsection{Instrumentation}
\label{sec:instr}

The first step of our evaluation is the instrumentation of the Bitnami
applications. Bitnami applications are distributed as self-contained virtual
machine (VM) images. \tool{} first extracts the virtual disk from the VM
image, assigns the disk local mount point, and creates a folder to store
program traces. Then, \tool{} edits the PHP interpreter configuration file
(i.e., \texttt{php.ini}) to enable Xdebug---a PHP extension that generates
function call tree files---and to change the default Xdebug settings
parameters\footnote{\tool{} requires the collection of full function variable
name and content, function return values, and a computer readable trace file
format. These are disabled by default. For more details, please refer
to~\cite{xdebug}.}. Finally,
\tool{} adds a system user and enables the OpenSSH server for the remote
access to retrieve call tree files.

After the instrumentation, \tool{} imports the VM image in the Virtual Box
hypervisor. It boots the VM and takes a snapshot. This snapshot will be the
starting point for the rest of the analysis.

\subsection{User Actions Input Trace}
\label{sec:useract}

We captured user actions traces using Selenium IDE~\cite{selenium}, a plugin
for Firefox. For each category of web application, we used two user roles:
regular user (e.g., customer for eCommerce applications) and administrator.
For each role, we registered user actions for a selection of web application
workflows. We focused on workflows that are common to all categories, such
as user sign-up and credential update, and workflows which are specific to a
category, e.g., invoice creation for accounting web applications.

\tool{} uses user actions traces both to generate dynamic traces and to test
the web application against \acsrf{}. In the first case, \tool{} replays all
user actions (See \Cref{sec:dyntrgen}). In the second case,
\tool{} replays only user login actions to update the HTTP request's session
cookie (See \Cref{sec:texec}). To distinguish user login actions from the
rest, we use the trace factoring functionality of Selenium IDE. More
specifically, we captured input traces as follows:

\begin{itemize}
  
  \item \textit{New workflow and no traces for a role}: We use Selenium IDE to
  capture the entire sequence of user actions of the workflow. Then, we factor
  actions in two sub-traces: one contains user login actions and the other
  contains workflow-specific actions. Each sub-trace is stored in its own
  file;

  \item \textit{New workflow and a trace for the user exists}: We import
  user login actions in Selenium IDE and then capture the new
  workflow-specific user actions;

  \item \textit{Same workflow but new user}: We duplicate the existing
  trace files, and replace credentials in the user login trace file. As traces
  are plain-text files, we use a script to find and replace user credentials.

\end{itemize}

The number of workflows (WFs) per web application is shown in
\Cref{tab:various}. The number varies according to availability of
off-the-shelf functionalities and the types of roles.

\subsection{Dynamic Traces Generation}
\label{sec:dyntrgen}
To generate dynamic traces, \tool{} replays user actions against an
instrumented VM. Action replaying is done step-by-step using Selenese Runner
Java (SRJ)~\cite{selenium}, an interpreter of Selenium user actions, that
controls a headless Firefox. The resulting requests are sent to an HTTP
proxy that forwards them one-by-one to the server. When the rendering process
of the browser is finished, SRJ signals that all statically referenced
external resources are retrieved (e.g., images, CSS). Then, \tool{} waits for
4 seconds (configurable) to honor any JavaScript asynchronous requests. After
that, no more requests are accepted, and the next action is fired. The first
request that entered the queue is associated to the fired user action. The
association is used during the model construction to establish causality.
Images and CSS are not likely to change the state and \tool{} does not include
them in the network trace. \tool{} uses a customizable list of MIME-types and
file extensions to exclude these resources.

Throughout the replaying of user actions, whenever \tool{} receives an HTTP
response, it accesses the VM to retrieve the generated PHP function call tree
and session data. The call tree file is associated to the request. This
association is used during the model construction to establish causality.
Finally, the call tree files are then processed to extract the MySQL queries
executed by the web application.

\subsection{Performance}

\begin{table}
    \footnotesize
    \centering
    \begin{tabular}{l  r | r |r r r| r }
        \toprule
       
        Web Apps      & WFs &  Tr. Gen. & Mod. Gen. & Nodes     & Edges      & Test \\
        \midrule
        AbanteCart    &  10 & 212s &  1,446s & 1,689,083 & 2,174,622 & 142s\\
        Horde         &   3 & 177s &    218s &    23,395 &    30,920 & 153s\\
        IN            &  11 & 152s &    215s &    97,465 &   123,419 &  82s\\
        Mautic        &   6 & 176s &    485s &   191,038 &   237,036 & 196s\\
        MyBB          &  12 & 214s &    261s &    96,766 &   119,270 & 183s\\
        OpenCart      &   8 & 179s &    312s &   160,401 &   224,351 & 123s\\
        Oxid          &  14 & 163s &    372s &   484,651 &   611,986 & 333s\\
        Prestashop    &  13 & 296s &    396s &   214,369 &   273,865 & 283s\\
        SI            &   9 & 128s &    170s &    34,248 &    44,983 &  31s\\
        SMF           &   7 & 134s &    159s &    61,738 &    78,893 & 493s\\
        \bottomrule

    \end{tabular}
    \caption{Execution time of \tool{}.}
    \label{tab:various}
\end{table}

In our assessment we used two computers. To generate traces and test for
execution, we used a workstation with an Intel i5-4690 CPU, an SSD disk and 32
GB of RAM. The workstation hosted a VirtualBox hypervisor that \tool{} used to
deploy Bitnami application containers. To generate our graph, we used a
workstation with an Intel i7-4600U CPU, an SSD disk and 12 GB RAM. We used a
single instance of Neo4j to handle property graphs of all applications with a
total of three million nodes and four million edges.

Overall, \tool{} took about \empirical{13} minutes to produce the output
report for a single web application (see \Cref{tab:various}). About 50\% of
the execution time is spent to generate traces and testing, which are largely
influenced by the web application behavior. For example, the first time that a
Prestashop webpage is requested, it creates a cache for frequently requested
resources. As we reset the virtual machine to the initial state, \tool{} waits
for Prestashop to re-create the local cache. Finally, model generation took in
average 7 minutes per web application. The execution of queries takes less
then 60s.

\subsection{Detection of \acsrf{}}

\tool{} discovered \empirical{29} security-relevant state-changing requests.
\empirical{17} of these tests detected a vulnerability in four web
applications: AbanteCart, Mautic, OpenCart, and Simple Invoices. The remaining
\empirical{12} requests did not detect vulnerabilities. We present attacks in
\Cref{sec:results}. 

\begin{table}
    \footnotesize
    \centering
    \begin{tabular}{l  r r | r | r}
        \toprule
        Web Apps.             & Reqs & SC Reqs & \multicolumn{2}{c}{Rel. SC Reqs$^{(*)}$}\\
        \midrule
        AbanteCart            &  335 &    335 &     8 & \textit{-98\%}\\
        Horde                 &   21 &     21 &     3 & \textit{-86\%}\\
        IN                    &  103 &    103 &    11 & \textit{-89\%}\\
        Mautic                &   58 &     21 &     8 & \textit{-62\%}\\
        MyBB                  &  104 &    104 &    21 & \textit{-80\%}\\
        OpenCart              &  117 &    117 &    11 & \textit{-91\%}\\
        Oxid                  &  165 &    165 &    10 & \textit{-94\%}\\
        Prestashop            &  267 &    195 &    16 & \textit{-92\%}\\
        SI                    &   92 &      7 &     7 & \textit{  0\%}\\
        SMF                   &  118 &    118 &    69 & \textit{-42\%}\\
        \midrule 
        Total                 &1,380 &  1,186 &   164 & \textit{-86\%}\\
        \bottomrule
        \multicolumn{5}{l}{\scriptsize\textit{* descrease \% from SC Reqs}}\\
    \end{tabular}
    \caption{Analysis results for the identification of relevant
    state-changing (SC) requests.}
    \label{tab:sc_ops}
\end{table}

\vspace{4pt}\noindent\textbf{\acsrf{} Candidates}---\Cref{tab:sc_ops} shows the number of
state-changing operations (column ``SC Reqs'') compared with the total number
of operations (column ``Reqs''). Results are aggregated by web application.
Almost all operations change the state. However, not all of these operations
are necessarily relevant for the security analysis. For example, some
operations may merely log user activities or be used to manage user sessions.
Thus, within a workflow, these operations most likely reoccur multiple times.
\Cref{tab:sc_ops} (column ``Rel. Reqs'') shows the total number of relevant
state-changing operations. The number of relevant operations decreased
considerably, i.e., on average by \empirical{-86\%}, from \empirical{1,186} to
\empirical{164}. The decrease is more evident in applications like AbanteCart,
where the number of operations decreased by \empirical{98\%} (from
\empirical{335} to \empirical{8}), whereas in other cases like Simple
Invoice, the number remained unchanged.

We manually inspected SQL queries that were excluded to assess the accuracy of
our heuristic. The total number of abstract SQL queries of our testbed is
\empirical{704}, of which \empirical{285} are considered not relevant. All
these queries are used to perform one of the following operations: session
management (e.g., creating a user session and refreshing of session token
validity), logging URL access, tracking user activity, and cache management
(e.g., MyBB stores entire CSS files in the DB). As these queries are not
relevant for our analysis, we conclude that our heuristic is accurate.

\vspace{4pt}\noindent\textbf{Security Tokens}---\tool{} identified \empirical{356}
variables of HTTP requests. \empirical{248} of them are discarded as they
are cookies (\empirical{192} variables), boundary markers of the multi-part
form data encoding (29 variables), and parameter names used
with timestamps\footnote{This technique is often used to bypass browser
caching mechanisms} (27 variables). These parameters cannot successfully
protect against \acsrf{} vulnerabilities. The remaining \empirical{108}
variables may be anti-CSRF tokens and are used by \empirical{53} operations
out of \empirical{164}. The remaining \empirical{111} state-changing
operations are not protected.

\vspace{4pt}\noindent\textbf{Security Testing}---\Cref{tab:tests} shows the total number
of tests that were generated for each approach. In total, we executed
\empirical{111} tests for unprotected operations and \empirical{108} for
protected ones. \tool{} monitored the test execution by using the sensors
installed during the instrumentation of the application container. In total,
\empirical{29} tests were successful and discovered severe vulnerabilities. We
discuss these results in detail in \Cref{sec:results}. The remaining
\empirical{190} tests failed. The majority of failed tests among the protected
operations are caused by the presence of an anti-CSRF token. In
\Cref{sec:results}, we present an in-depth discussion of the use of this
token.
The remaining failed tests (including several unprotected operations) are
caused by multi-step workflows in which the tested HTTP request depends on
another request that is not part of the test. We leave the study of
dependencies between requests as a future research direction.

\begin{table}
    \footnotesize
    \centering
    \begin{tabular}{l   r   r r | r | r   r r | r}
        \toprule
        Web Apps.         &  \multicolumn{4}{c}{Protected}  & \multicolumn{4}{c}{Unprotected} \\
        \cmidrule(lr){2-5}
        \cmidrule(lr){6-9}
                                &   TCs & Fail. & Succ. &     Expl. & TCs$^{(*)}$ & Fail. & Succ. & Expl.\\
        \midrule
        \textbf{AbanteCart}     &     3 &    2  &     1 &\textbf{1} &           5 &      2 &    3 &\textbf{2}\\
        Horde                   &     3 &    3  &     - &         - &           - &      - &    - &         -\\
        IN                      &    12 &   12  &     - &         - &           - &      - &    - &         -\\
        \textbf{Mautic}         &    19 &   17  &     2 &\textbf{2} &           - &      - &    - &         -\\
        MyBB                    &     1 &    1  &     - &         - &          20 &      9 &   11 &         -\\
        \textbf{OpenCart}       &     2 &    1  &     1 &\textbf{1} &           9 &      5 &    4 &\textbf{4}\\
        Oxid                    &    33 &   33  &     - &         - &           - &      - &    - &         -\\
        Prestashop              &     7 &    7  &     - &         - &          11 &     11 &    - &         -\\
        \textbf{SI}             &     - &    -  &     - &         - &           7 &      - &    7 &\textbf{7}\\
        SFM                     &    20 &   20  &     - &         - &          47 &     47 &    - &         -\\
        \bottomrule
        \multicolumn{6}{l}{\scriptsize\textit{* one TC for each unprotected operation}}\\
    \end{tabular}
    \caption{Generation and assessment of test cases. \textit{TCs=nos. of testcases, Fail./Succ.=nos.
    of un/successful tests, and Expl.=nos. of tests that exploited an \acsrf{}
    vulnerability} }
    \label{tab:tests}
\end{table}

\section{Results}
\label{sec:results}

We now detail the vulnerabilities that \tool{} discovered in
the four vulnerable web applications. We also discuss tests that
discovered state transitions that cannot be exploited in a \acsrf{} attack.

\subsection{Exploitable Vulnerabilities}
\label{sec:attacks}

Four web applications of our testbed are vulnerable to \acsrf{} attacks. The
severity of this vulnerability ranges from very high, i.e., customer account
takeover, website takeover, and database deletion, to low, i.e., adding items
into a shopping cart. These vulnerabilities can potentially affect millions of
websites. For example, according to Pellegrino et
al.~\cite{DBLP:conf/ndss/PellegrinoB14}, OpenCart is used by at least nine
million websites whereas AbanteCart is used by 21 thousand websites. We
responsibly disclosed these vulnerabilities to the developers. In this
section, we present a comprehensive overview of our findings and a detailed
description of the most severe issues.

\subsubsection{Overview of all Vulnerabilities} In summary, we 
discovered the following vulnerable operations:

\vspace{4pt}\noindent\textbf{AbanteCart}---An attacker can (i) take over a
customer's user account and (ii) add or modify the shipping address.
Developers have already fixed this vulnerability.

\vspace{4pt}\noindent\textbf{OpenCart}---An attacker can (i) take over a
customer's user account, (ii) add or modify the shipping address, and (iii)
add items to a customer's shopping cart\footnote{This vulnerability was also
found and reported by a third party in independent and parallel research.}.

\vspace{4pt}\noindent\textbf{Mautic}---An attacker can (i) delete a marketing
campaign (part of the core logic of the web application), and (ii) delete
recipients from a marketing campaign. Developers of Mautic were unresponsive
and we requested and obtained a CVE entry
(CVE-2017-8874).

\vspace{4pt}\noindent\textbf{Simple Invoices}---An attacker can (i) create new
website administrators and customers, (ii) enable payment methods, (iii)
create new invoices, and (iv) change taxation parameters. Developers of Simple
Invoices acknowledged the presence of the flaw, but they were not working on a
patch yet. Accordingly, to protect SI users, we requested and obtained a CVE
entry (CVE-2017-8930).

\subsubsection{Attack \#1: Account Takeover with
AbanteCart and OpenCart} The vulnerable state-changing operations of both
web applications are not protected by anti-CSRF tokens.

The attack against OpenCart exploits two \acsrf{} vulnerabilities in the 
operations to (i) change the user email address
and (ii) to update user passwords. When changing this
security-sensitive information, OpenCart neither uses anti-CSRF tokens, nor
requires users to provide their current password. As a result, an attacker can
use \acsrf{} to reset both
email and password to hijack an account.

The attack against AbanteCart exploits the \acsrf{} vulnerability in the
operation to change user data (e.g., email address, first and last name). As
opposed to OpenCart, AbanteCart does not use the email address as username.
However, it permits recovering usernames and resetting user passwords via the
``forgot username'' and ``forgot password'' features. To reset the username,
AbanteCart asks for an email address and the last name of the customer, then
sends the username in an email. As the attacker can change the email and last name
with an \acsrf{} attack, she can successfully retrieve the username. The
``forgot password'' requires the username and the email address. As the
attacker possesses both, she receives a link to reset the password
via email.

\subsubsection{Attack \#2: Database Corruption in Mautic}

Our tests discovered two \acsrf{} vulnerabilities in Mautic which allow an
attacker to compromise the core functionalities of the software. Mautic is a
marketing automation web application which allows users to create email
marketing campaigns and to manage the contacts of the campaign. Our tests
discovered \acsrf{} vulnerabilities in these two operations in which an
attacker can delete a specific campaign or a contact. The identifier used to
refer to both campaigns and contacts is an incremental integer number. An
attacker can either compromise specific campaigns by deleting them or by
deleting users, or can delete all existing campaigns and contacts.

\subsubsection{Attack \#3: Web Application Takeover with Simple Invoices}

Our analysis discovered that seven state-changing operations in Simple
Invoices are not protected by any session-unique or user-unique data value. In
total, six workflows are vulnerable to \acsrf{} vulnerabilities. These
workflows are: creation of a new website administrator, creation of a new
customer account, enabling payment methods (e.g., PayPal), adding a new invoice to
the database, and changing both global and invoice tax rates.

\subsection{Non-Exploitable Tests}

\empirical{11} tests caused a change of state in MyBB. The operations
under test were privileged operations performed by the website administrator.
While the tests were successfully executed, they cannot be exploited by an
attacker. MyBB uses a secret user-unique API key which authenticates the user
when performing state-changing requests. If the key is valid, then the
operation is executed. While for regular users, in our model this key is
correctly labeled unique per user, for the administrator, the key is labeled
constant. In our analysis, we used traces from a single administrator user, as
MyBB has no concept of multiple administrator accounts.  Thus, all these
traces contained the same key, causing our type inference algorithm to infer
the constant type. Accordingly, the key is included in our tests. The
server-side program verifies that the key belongs to the administrator and
executes the requested operation.

\section{Analysis}

Despite its popularity and severity, our results show that the risk posed by
\acsrf{} vulnerabilities is overlooked or even misunderstood. An analysis of 
our results exposes three distinct classes of developer awareness---complete, 
partial and nonexisting: 

\vspace{4pt}\noindent\textbf{Complete Awareness}---At one end of the
awareness spectrum, we have full awareness, in which developers deploy \acsrf{}
countermeasures for \emph{all} state-changing operations. Examples of this
group are Horde, Oxid, and Prestashop. For example, in the case of Oxid,
all \empirical{33} tests failed when omitting an anti-CSRF token.

\vspace{4pt}\noindent\textbf{Unawareness}---At the other end, we have complete
unawareness. Developers may still not be aware of \acsrf{} nor of the security
implications of successful exploitations. As a result, developers may leave
state-changing operations unguarded. Simple Invoices is an example of such a
case, in which all state-changing operations are vulnerable to \acsrf{}
attacks.

\vspace{4pt}\noindent\textbf{Partial Unawareness}---We observed two interesting cases in
which protections are deployed in a selective manner. From our testbed, we can
distinguish two clear cases.

\vspace{2pt}\noindent\textit{Role-based Protections}: Examples for this case are OpenCart
and AbanteCart, which treat regular users and administrators differently. Our tests
showed that administrator operations are protected by anti-CSRF tokens.
Omitting these tokens results in rejected state-changing operations. This shows
that developers are aware of the security risks and that they deployed
adequate countermeasures. However, user operations are not equally protected.
As we have seen, even critical operations, such as password change, are
exposed to severe attacks leading to customer account takeover. We speculate
that this may be the result of an inadequate or incomplete risk analysis and
threat modeling during the design phase.

\vspace{2pt}\noindent\textit{Operation-based Protections}: As opposed to the previous case, the
distinction is not based on the role of the user, but on the type of
operation. In general, web applications offer operations to create, delete,
and update elements in a database. Elements can be anything including users,
contacts, and products. In Mautic, we observed that creation and updating are
guarded by anti-CSRF tokens. \tool{} verified that when a token is omitted, a
test fails. Similarly for the cases of AbanteCart and OpenCart, this behavior
shows that the developers may be aware of the security risks. However,
deletion operations are not protected, allowing attackers to compromise the
database. In contrast to role-based protections, this may not be caused by
inadequate threat modeling. We believe that developers just overlooked this
operation.

\section{Discussion and Future Work}

\vspace{4pt}\noindent\textbf{Scalability of the Model}---Our assessment showed that a
modern workstation can efficiently handle a single graph database instance
with three million nodes. We believe that this would be an average use case of our
tool. However, property graphs can scale to hundreds of millions of
nodes~\cite{backes2017}. In these scenarios, \tool{} can also be run on
servers, exploiting the availability of additional hardware resources.

\vspace{4pt}\noindent\textbf{Performance}---The main bottleneck of our approach is the
interaction with a running web application. In our experiments, we used one
virtual machine at a time, but, we plan to improve performance by spawning
parallel, multiple virtual machine instances of the same web application.

\vspace{4pt}\noindent\textbf{Generality of the Approach}---Our evaluation was conducted on
PHP-based web applications using a MySQL database. While these are popular
among web developers, web applications can use different SQL databases or can
be written in other programming languages. The modeling framework is
independent from the programming language. However, instrumentation and
sensors may require new connectors in order to acquire traces.

\vspace{4pt}\noindent\textbf{Detection Power}---\tool{} was conceived to target \acsrf{}.
However, as for CSRF, other classes of severe vulnerabilities have been
neglected by the security community, e.g., session management issues and race
conditions. The lowest common denominator of these classes is that they are
much more complex to detect when compared to XSS and SQLi. The detection
of these classes require learning in-depth behaviors of a program and
synthesizing the relevant aspects in models. From this point of view, our
modeling paradigm has to be seen as an initial effort toward this long-term
goal. \tool{} provides a unified representation for artifacts and models used
in dynamic analysis, and furthermore, it provides a semantic of the
relationships between them. However, our representation may not be sufficient
to capture relevant aspects for the detection of other classes of
vulnerabilities.

\section{Related Work}
\label{sec:related}

To the best of our knowledge, this is the first work proposing a technique for
the detection of \acsrf{} vulnerabilities. Existing work focused
mainly on defense techniques, proposing new HTTP headers
(See, e.g.,~\cite{Barth:2008:RDC:1455770.1455782, Mao2009,
Johns_requestrodeo:client, JovanovicKK06}) and new CSRF-based
attacks (e.g.,~\cite{Sudhodanan2017}). As opposed to these works,
\tool{} does not protect from exploitation, but it allows discovery of CSRF
during the testing phase of the development of web applications.

\vspace{4pt}\noindent\textbf{Property Graphs and Vulnerability Detection}---Our approach
relies on graph databases for the representation and composition of models.
Similar to our idea, Yamaguchi et
al.~\cite{Yamaguchi:2014:MDV:2650286.2650771} and Backes et
al.~\cite{backes2017} combined different code representations in a property
graph. While these works focused on static source code representations, we
model dynamic behaviors of the application. Furthermore, these works,
similarly to others in the area of web security, focused on input validation
vulnerabilities. In contrast, our work presented a technique to discover
\acsrf{}.

\vspace{4pt}\noindent\textbf{Dynamic Analysis}---Research on dynamic analysis has been
very active over the last decade, proposing new techniques and tools to detect
a variety of vulnerabilities. For example, unsupervised web application
scanners are very popular tools routinely used to detect vulnerabilities in
web applications. Starting from a URL, a web application scanner crawls a web
application and then, for each discovered input, it probes the application
with crafted input strings. There are plenty of commercial and non-commercial
scanners, including tools proposed by the research
community~\cite{Kals:2006:SWV:1135777.1135817, Huang:2005:TFW:1090583.1090587,
stateaware, Pellegrino2015, Mesbah:2012:CAW:2109205.2109208}). While web
scanners are effective in the detection of XSS and SQLi, they still perform
poorly or even fail in the detection of more sophisticated vulnerabilities,
including \acsrf{} vulnerabilities~\cite{johnny, sota_scanners}. Compared to
web scanners, \tool{} does not include a crawler component. Crawlers use
breadth- or depth-first algorithms which are not adequate to reach
security-relevant state-changing requests. As opposed to this technique,
\tool{}---similarly to other dynamic approaches (See,
e.g.,~\cite{DBLP:conf/ndss/PellegrinoB14,
Mcallister:2008:LUI:1433006.1433021})---follows a different approach in which
input traces are used to explore in depth the functionalities of web
applications. Other approaches have been proposed in order to address more
complex flaws, e.g., user authentication (see, e.g., ~\cite{authscan,
Zhou:2014:SAT:2671225.2671257}), and logic vulnerabilities
(e.g.,~\cite{DBLP:conf/ndss/PellegrinoB14}), often combining model inference
with dynamic testing. These approaches analyze components and functionalities
that are specific to the vulnerability being targeted, thus making them
inherently limited in the ability to reason about the presence of CSRF
vulnerabilities.

\vspace{4pt}\noindent\textbf{Static Analysis}---Static program analysis has been used to
detect several classes of vulnerabilities, e.g., input validation
vulnerabilities~\cite{Jovanovic:2006:PSA:1130235.1130378, webssari,
Dahse_stored, backes2017}, authorization vulnerabilities~\cite{MACE}, and
logic flaws~\cite{whitebox_logic}. Similarly as for dynamic techniques, none
of the existing approaches target CSRF vulnerabilities. Second, more and more
web applications tend to use programming languages and coding patterns, e.g.,
runtime second-order function calls~\cite{DBLP:conf/issta/HillsKV13,
DBLP:conf/wcre/000115} and SQL query
construction~\cite{DBLP:conf/wcre/AndersonH17}, that are hard to treat
statically. Static analyzers often address these shortcomings by calculating
over- or under-approximations that can cause high rates of false
positives~\cite{backes2017}. In these scenarios, dynamic techniques such as
\tool{} are a valid alternative; however, existing approaches lack the
sophistication to detect CSRF.

\section{Conclusion}

We presented \tool{}, to the best of our knowledge the first security testing
framework that can detect \acsrf{} vulnerabilities. At the core of \tool{} is
a new modeling paradigm based on property graphs that defines (i) searchable
model components to represent multiple aspects of web applications, and (ii) a
query language that allows expression of suspicious or vulnerable behaviors.
Our experiments detected \empirical{14} severe \acsrf{} vulnerabilities
affecting four web applications that can be used to take over websites, or
user accounts, and compromise database integrity. Finally, we assessed the
current awareness level of the \acsrf{} vulnerabilities and showed alarming
behaviors in which security-sensitive operations are protected in a selective
manner. This work has successfully demonstrated the capabilities of our
paradigm, which comprehensively captures non-trivial, cross-tier aspects of
modern web applications. In the near future, we intend to leverage the
opportunities provided by our paradigm and extend the approach towards
additional vulnerability classes.

\section*{Acknowledgments}

We would like to thank the anonymous reviewers for their valuable feedback and
our shepherd Adam Doup\'{e} for his support in addressing reviewers' comments.
We would like also to thank Benny Rolle and Florian Loch for their
contribution to the development of \tool{}. This work was supported by the
German Federal Ministry of Education and Research (BMBF) through funding for
the Center for IT-Security, Privacy and Accountability (CISPA) (FKZ:
16KIS0345, 16KIS0656), the CISPA-Stanford Center for Cybersecurity (FKZ:
13N1S0762), and the project BOB (FKZ: 13N13250).
 
\bibliographystyle{ACM-Reference-Format}
\bibliography{literature,rfc}

\end{document}